\documentclass[twocolumn]{aastex63}
\usepackage[utf8]{inputenc}
\usepackage{comment}
\usepackage{amsmath}
\usepackage{graphicx}

\newcommand{\zp}{$z_\textrm{phot}$}

\begin{document}
\title[Implications of a Temperature Dependent IMF]{Implications of a Temperature Dependent IMF I: Photometric Template Fitting} 

 
\author[0000-0002-5460-6126]{Albert Sneppen}
\affiliation{Niels Bohr Institute, University of Copenhagen, Lyngbyvej 2, K\o benhavn \O~2100, Denmark}
\affiliation{Cosmic Dawn Center (DAWN)}

\author[0000-0003-3780-6801]{Charles L. Steinhardt}
\affiliation{Cosmic Dawn Center (DAWN)}
\affiliation{Niels Bohr Institute, University of Copenhagen, Lyngbyvej 2, K\o benhavn \O~2100, Denmark}

\author{Hagan Hensley}
\affiliation{California Institute of Technology, 1200 E. California Blvd., Pasadena, CA 91125, USA}
\affiliation{Cosmic Dawn Center (DAWN)}

\author{Adam S. Jermyn}
\affiliation{Center for Computational Astrophysics, Flatiron Institute, New York, NY 10010, USA}

\author{Basel Mostafa}
\affiliation{California Institute of Technology, 1200 E. California Blvd., Pasadena, CA 91125, USA}
\affiliation{Cosmic Dawn Center (DAWN)}

\author[0000-0003-1614-196X]{John R. Weaver}
\affiliation{Niels Bohr Institute, University of Copenhagen, Lyngbyvej 2, K\o benhavn \O~2100, Denmark}
\affiliation{Cosmic Dawn Center (DAWN)}


\begin{abstract}
A universal stellar initial mass function (IMF) should not be expected from theoretical models of star formation, but little conclusive observational evidence for a variable IMF has been uncovered. In this paper, a parameterization of the IMF is introduced into photometric template fitting of the COSMOS2015 catalog. The resulting best-fit templates suggest systematic variations in the IMF, with most galaxies exhibiting top-heavier stellar populations than in the Milky Way.  At fixed redshift, only a small range of IMFs are found, with the typical IMF becoming progressively top-heavier with increasing redshift. Additionally, subpopulations of ULIRGs, quiescent- and star-forming galaxies are compared with predictions of stellar population feedback and show clear qualitative similarities to the evolution of dust temperatures.   \newline 
\end{abstract}

\section{Introduction}
\label{sec:intro}
The total luminosity of a galaxy is dominated by the most massive stars in its stellar population, yet these individual stars cannot be directly characterized. As a result, nearly everything that has been inferred about the properties of high-redshift galaxies relies on an assumed stellar initial mass function (IMF) \citep{Bonnell2007,Conroy2009}.  Several key quantities depend particularly strongly on the shape of the assumed IMF.  For example, because the stellar mass is dominated by low-mass stars, inferred stellar masses and star formation rates (SFRs) are highly sensitive to the ratio of high-mass to low-mass stars in the IMF.  In photometric template fitting, the most common technique used in large, high-redshift surveys, the inferred extinction, metallicity, and other properties are also covariant with the assumed IMF (cf. \citet{Brammer2008}).  Thus, it is essential not merely to assume an IMF, but to assume the correct one.  

On a fundamental level, star formation involves a competition between gravity and processes which provide support against collapse, most notably thermal pressure.  Therefore, the IMF must be sensitive to anything which affects that balance.  This should include a wide range of properties that differ between galaxies, including the central potential, existing stellar mass, star formation history, supernova rate, cosmic ray density and galactic magnetic fields, metallicity, dust density and composition, AGN activity, and the environment and merger history.  All of these are known to vary both between different galaxies at fixed redshift and across different redshifts (cf. \citet{Laigle2016}).  Therefore, it should be expected that the IMF is not universal, but rather differs between galaxies, and between different epochs for the same galaxy.  In particular, the IMF should depend upon the gas temperature of star-forming clouds, with higher-temperature regions producing higher average stellar masses \citep{Lynden-Bell1976,Jermyn2018}.  Since observations of dust even at moderate redshifts find an increase in temperature towards higher redshift \citep{PPP2010,Casey2012,Magnelli2014,Magdis2017}, this implies that galactic IMFs should systematically vary with redshift, and thus the assumed IMF must do the same.  

However, at present nearly all studies have assumed a universal IMF, choosing one of several estimated Galactic IMFs and applying that to every galaxy at all redshifts.  In part, this is out of necessity.  It has not yet been possible to directly observe the IMF outside of the Galaxy and local satellites.  Indeed, even these local measurements have produced several proposed Galactic IMFs \citep{Salpeter1955,Kroupa2001,Chabrier2003}.  Furthermore, inferred properties such as stellar mass and SFR are known to be highly sensitive even to which of the proposed Galactic IMFs is assumed (cf. \citet{Speagle2014}).  Attempts to constrain extragalactic IMFs, or even to determine whether the same IMF fits all galaxies, have produced little consensus \citep{Cenarro2003,Treu2010,Cappellari2012}.  Although results at times have suggested both top-heavier \citep{Conroy2012} and bottom-heavier \citep{Martin-Navarro2015} IMFs at high redshift, there is not yet conclusive evidence that any specific galaxy has a non-Galactic IMF.  

This limited success has primarily been caused by two distinct problems.  First, there are significant degeneracies between the IMF and extinction, metallicity, star-formation history, and the age of the stellar population.  Thus, it is likely impossible to determine the entire shape of the IMF from high-redshift observations alone.  Second, theory has not yet produced consensus predictions for the functional forms that IMFs might take at higher redshift.

In recent work, \cite{Jermyn2018} proposed a temperature-dependent family of IMFs to account for temperature variation.  This family of IMFs, combined with cosmic ray feedback, was used to predict a temperature history for the star-forming "main sequence" \citep{Steinhardt2020}.  Because this family of IMFs only introduces one additional parameter, $T_{IMF}$, it is possible to fit photometric catalogs with the entire family of IMFs, producing a best-fit IMF in addition to the other fit parameters.  

In \S~\ref{sec:method}, we describe the introduction of this additional parameter into EAZY \citep{Brammer2008}, a leading photometric template fitting code.  The modified fitting routines are then applied to the COSMOS2015 catalog \citep{Laigle2016}, the largest current multi-wavelength photometric catalog, in \S~\ref{sec:results}.  At most redshifts, galaxies are best fit with a higher $T_{IMF}$ than in the Milky Way, corresponding to a top-heavier IMF.  In \S~\ref{sec:tests}, we discuss the results of various tests to determine whether these fits are truly constraining the IMF, rather than other parameters with which the IMF is covariant or even fully degenerate.  Finally, several key implications of these results are discussed in \S~\ref{sec:discussion}. This constitutes Paper I in a series of three related papers, with the remaining implications of these new measurements reserved for Papers II and III.  In Paper II, the star-forming galaxies and the star-forming main sequence are discussed and Paper III focuses on quiescent galaxies and quenching.

The analysis presented here assumes a flat $\Lambda$CDM cosmology with $(h, \Omega_m, \Omega_\Lambda) = (0.674 ,0.315, 0.685)$ \citep{Planck2018} throughout.

\section{Methodology}
\label{sec:method}
This section describes the parameterization of a family of IMFs and the dataset and fitting techniques used to find the best-fit IMF for galaxies in the COSMOS2015 catalog.

\subsection{The COSMOS2015 Dataset}
\label{subsec:catalog}

In this work, fitting is performed on the COSMOS2015 catalog \citep{COSMOS2015}, the largest and most comprehensive multi-wavelength catalog available.  COSMOS2015 includes 10 broad-band filters spanning NUV (from GALEX), u (from CFHT/MegaCam) B, V, r, i, z++, Y, J, H, and IRAC channels 1 and 2, two narrow band filters (NB711 and NB816), and 12 intermediate bands (IA427, IA464, IA484, IA505, IA527, IA574, IA624, IA679, IA709, IA738, IA767 and IA827) from the COSMOS-20 survey.  Most objects are not covered by every filter, with further elaborations on the catalog detailed in \citet{COSMOS2015}. 518,403 galaxies in the COSMOS2015 catalog are not flagged as saturated or within bad regions. However, to properly deconvolve the IMF reasonable uncertainties are required in the photometric fitting code,  {\textbf so a V-band SNR-cut of 10 is used, with 139,535 galaxies passing this cut.  In Paper III, an alternative K-band cut is considered, finding that the key results are insensitive to the choice of band and merely require a sufficiently high SNR across the spectrum.}

\subsection{Temperature Dependence of the Stellar IMF}

The parameterizations used in this work are modified versions of the Kroupa IMF \citep{Kroupa2001},
\begin{equation}
    \frac{dN}{dm} \propto 
    \begin{cases}
       m^{-0.3} &\quad m < \tilde{m} \\
       m^{-1.3} &\quad \tilde{m} < m < \tilde{M} \\
       m^{-2.3} &\quad \tilde{M} < m, \\
    \end{cases}
    \label{kroupa_1}
\end{equation}
with $\tilde{m} = 0.08 M_{\odot}$ and $\tilde{M} = 0.5 M_{\odot}$ empirically determined for the Galactic IMF.  

Physically, the mass scale scales $\tilde{m}$ and $\tilde{M}$ are believed to be the minimum masses for fragmentation and collapse, respectively.  Both processes depend upon gas temperature.  Cloud collapse has long been modeled with a Jeans approximation \citep{Jeans1902}, and fragmentation has a similar dependence during adiabatic collapse \citep{Lynden-Bell1976,Bonnell2007}.  As a simplification, here it is also assumed that as the cloud collapses its temperature remains invariant so long as it is optically thin~\citep{Steinhardt2020}.

\begin{figure}[!ht]
\begin{center}
\includegraphics[width=.95\linewidth]{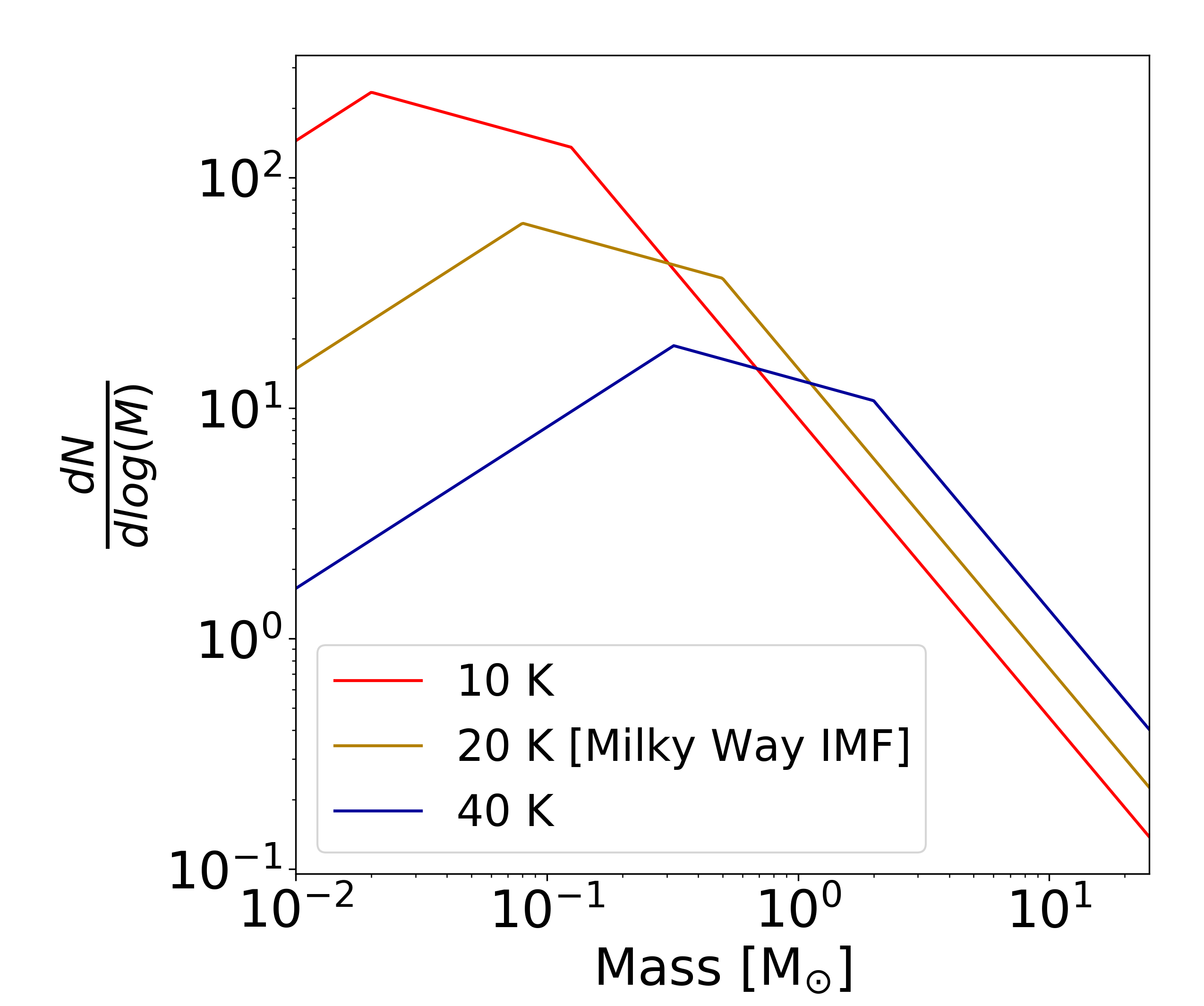}
\caption{A Kroupa initial mass function normalized by total mass with described temperature dependence at T = 10 K [red], 20 K [orange], and 40 K [blue]. Note that raising the temperature leads to a more top-heavy or more bottom-light distribution and lowering the temperature leads to a more bottom-heavy distribution.} 
\label{fig:imf}
\end{center}
\end{figure}

\citet{Jermyn2018} used these two processes to derive a temperature-dependent family of IMFs, 
\begin{equation}
    \frac{dN}{dm} \propto 
    \begin{cases}
       m^{-0.3} &\quad m < 0.08 M_{\odot} \cdot \bigg( \frac{T_g}{T_0} \bigg)^2 \\
       m^{-1.3} &\quad 0.08 M_{\odot} \cdot \bigg( \frac{T_g}{T_0} \bigg)^2 < m < 0.5 M_{\odot} \cdot \bigg( \frac{T_g}{T_0} \bigg)^2 \\
       m^{-2.3} &\quad 0.5 M_{\odot} \cdot \bigg( \frac{T_g}{T_0} \bigg)^2 < m, \\
    \end{cases}
    \label{eq:kroupa_2}
\end{equation}
where $T_g$ is the gas temperature in star-forming molecular clouds.  They argue that physically, both $\tilde{m}$ and $\tilde{M}$ scale as $T_g^2$, for independent reasons.  The typical Galactic star-forming cloud is at $T \sim 20$ K, which is set as the reference temperature $T_0$ \citep{Schnee2008}.  At $T_g = T_0 = 20$ K, this yields the standard Kroupa IMF.  At higher temperatures, this produces a top-heavier IMF, and a lower temperatures, a bottom-heavier one (see Fig. \ref{fig:imf}).

It should be noted that the specific dependence of $\tilde{m}$ on temperature requires solving the energy balance and opacity with a variety of different assumptions (e.g., opacity laws, the ambient temperature of surrounding medium, the angular rotation of the cloud, the ratio of turbulent pressure to total pressure, the strength of magnetic fields), resulting in a wide range of scaling relations, including $\tilde{m} \propto T$ \citep{imf-t}, $\tilde{m} \propto T^{3/2}$ \citep{Jeans}, $\tilde{m} \propto T^{2}$ \citep{Steinhardt2020}, and $\tilde{m} \propto T^{5/2}$ \citep{imf-t52}. Thus, some attention should be given to the interpretation of the actual best-fit top-heaviness of the IMF, i.e. $T_{IMF}$, which using the assumptions of~\citet{Jermyn2018} corresponds to the gas temperature, $T_{g}$.  For other temperature dependencies, the translation to $T_{g}$ is straight-forward (Table \ref{tab:tdependence}).  The IMF parameter will correctly describe the shape of the IMF regardless of the proper association with gas temperature.

\begin{table}[!ht]
\begin{center}
 \begin{tabular}{lll} 
 \hline
 Reference & Mass Scaling & $T_g - T_{IMF}$ Relationship \\ 
 \hline
 Hopkins 2012 & $\left( \frac{T}{T_0} \right)$  & $\frac{T_g}{20\textrm{ K}} = \left(\frac{T_{IMF}}{20\textrm{ K}} \right)^{2}$   \\
 Jeans 1902 & $\left( \frac{T}{T_0} \right)^{3/2}$ & $\frac{T_g}{20\textrm{ K}} = \left(\frac{T_{IMF}}{20\textrm{ K}}\right)^{4/3} $  \\
 Steinhardt 2020 & $\left( \frac{T}{T_0} \right)^2$ &  $T_g = T_{IMF}$  \\
 Chabrier 2014 & $\left( \frac{T}{T_0} \right)^{5/2}$ & $\frac{T_g}{20\textrm{ K}} = \left( \frac{T_{IMF}}{20\textrm{ K}} \right)^{4/5} $  \\
 \hline
\end{tabular}
\caption{The relationship between the IMF parameter used here, $T_{IMF}$, and gas temperature $T_g$ using different physical models for the temperature dependence of the IMF.}
\label{tab:tdependence}
\end{center}
\end{table}

One might expect that in practice, molecular clouds in a star-forming galaxy should exhibit a range of temperature, and that therefore the true IMF might be a composite of several different components.  Further, older stellar subpopulations might have formed under different conditions, and with a different IMF, than younger ones.  

In this work, it is instead assumed that every galaxy has a single IMF, and that the same IMF can be used for the entire stellar population.  This is done for two reasons. The first is practical; the IMF from Eq. (\ref{eq:kroupa_2}) only requires the introduction of a single additional parameter, $T_{IMF}$.  Thus, photometric catalogs with a limited number of bands can still provide meaningful constraints on the best-fit $T_{IMF}$.  This is similar to other approximations; for the same reason it is also standard in template fitting to fit the stellar population as having a single age, and to fit the galaxy with a single extinction, metallicity, etc., even though in practice galaxies are not monolithic.  The other is that physical mechanisms have been proposed which could create similar conditions in star-forming regions throughout a galaxy.  For example, \citet{Steinhardt2020} proposed a model in which the star-forming main sequence would be driven by feedback between a temperature-dependent IMF and cosmic ray feedback.  In this model, because cosmic rays can permeate the entire galaxy, star-forming regions would all have similar gas temperatures.  We evaluate the predictions of this model in \S~\ref{subsec:crpredictions}.

In principle, the methods developed in this work could also be used to test this assertion more directly.  In recent works, photometry of resolved portions of galaxies have been fit independently to determine whether the same best-fit properties are consistent throughout the same galaxy \citep{Sorba2018,Greener2020,ValeAsari2020,Fetherolf2020}, albeit with mixed results.  The same could be done with the new templates produced here.

\subsection{Introducing variable IMFs into photometric template fitting}
Throughout this work, fitting is performed using the photometric template fitting code Easy and Accurate \zp\ from Yale (EAZY \citealt{Brammer2008}; source code can be accessed at http://www.astro.yale.edu/eazy/), which has been shown to produce reliable photometric redshifts when compared against other software \citep{Dahlen2013, Hildebrandt2010}. 

EAZY fits photometry by searching for the closest reconstructed spectrum from a library of synthetic spectra.  Specifically, the process begins by constructing a larger basis of 560 synthetic templates are derived using the Flexible Stellar Population Synthesis code (FSPS; \citealt{fsps1}, \citealt{fsps2}) which span a variety of ages, metallicities, star-formation histories, dust attenuation and extinction curves (Table \ref{fsps_param}). These synthetic spectra follow FSPS convention and includes nebular continuum and line emission features derived from the Cloudy models \cite{Byler2017}. Then, a smaller basis of 12 theoretical spectra is constructed to span the larger basis of 560 spectra using non-negative matrix factorisation (NNMF). The NNMF reduction takes a large number of synthetic models and computes a reduced basis of templates which best reproduce synthetic broad-band photometric calibration catalogues (see \citet{Brammer2008} for a complete discussion).  Thus, the reduced basis represents the "principal-component” spectral templates. This analysis follows convention by utilizing the reduction coefficient already derived and utilized in the established EAZY templates \citep{Brammer2008}.  

The assumption is then that linear combinations of the reduced basis spans the entire range of real galaxy spectra, yet with few enough parameters that the best fit can be constrained from a small number of photometric bands.  EAZY then finds the best-fit linear combination of these 12 basis spectra to the observed photometry, while accounting for redshift and convolving with filter profiles.  Best-fit galaxy properties are then calculated as a (luminosity-)weighted average of the properties corresponding to each template.
 
\begin{table}[!ht]
\begin{center}
 \begin{tabular}{||  m{5cm} | >{\centering\arraybackslash}m{2cm} ||} 
 \hline
 Parameter & Range \\ 
 \hline\hline
 Age [Gyr] & 0.02-20 \\ 
 \hline
 $\tau$ e-folding time [Gyr] &  0.05 - 0.8  \\
 \hline
 Metallicity [$Z_{\odot}$] & 0.1 - 1.78  \\
 \hline
 Dust extinction, $A_V$ [mag] & 0.01 - 3  \\
 \hline
 Temperature [K] & 8 - 50  \\ [1ex] 
 \hline
\end{tabular}
\caption{Range of galactic parameters spanned in the initial 560 FSPS spectra.  These include the average age of the stellar population, characteristic time for star formation $\tau$, metallicity taken from the tabular Mist isochrones, and the dust extinction $A_v$ in magnitudes.}
\label{fsps_param}
\end{center}
\end{table}

In introducing an additional parameter, the goal is to alter this process as little as possible, so that previous experience using and interpreting EAZY fits will still be applicable.  Thus, the only novel part of this approach comes in the initial template definition.  The larger basis of 560 FSPS templates is calculated independently for every $T_{IMF}$ in a grid of temperatures.  Those are then reduced to a smaller, 12-template basis as before, and then used to find a best-fit linear combination to the photometry.  Comparing the residual $\sum \chi^2_n$ for each fit allows the determination of a single best fit across all choices of $T_{IMF}$.

\begin{figure}[!ht]
\begin{center}
\includegraphics[width=\linewidth]{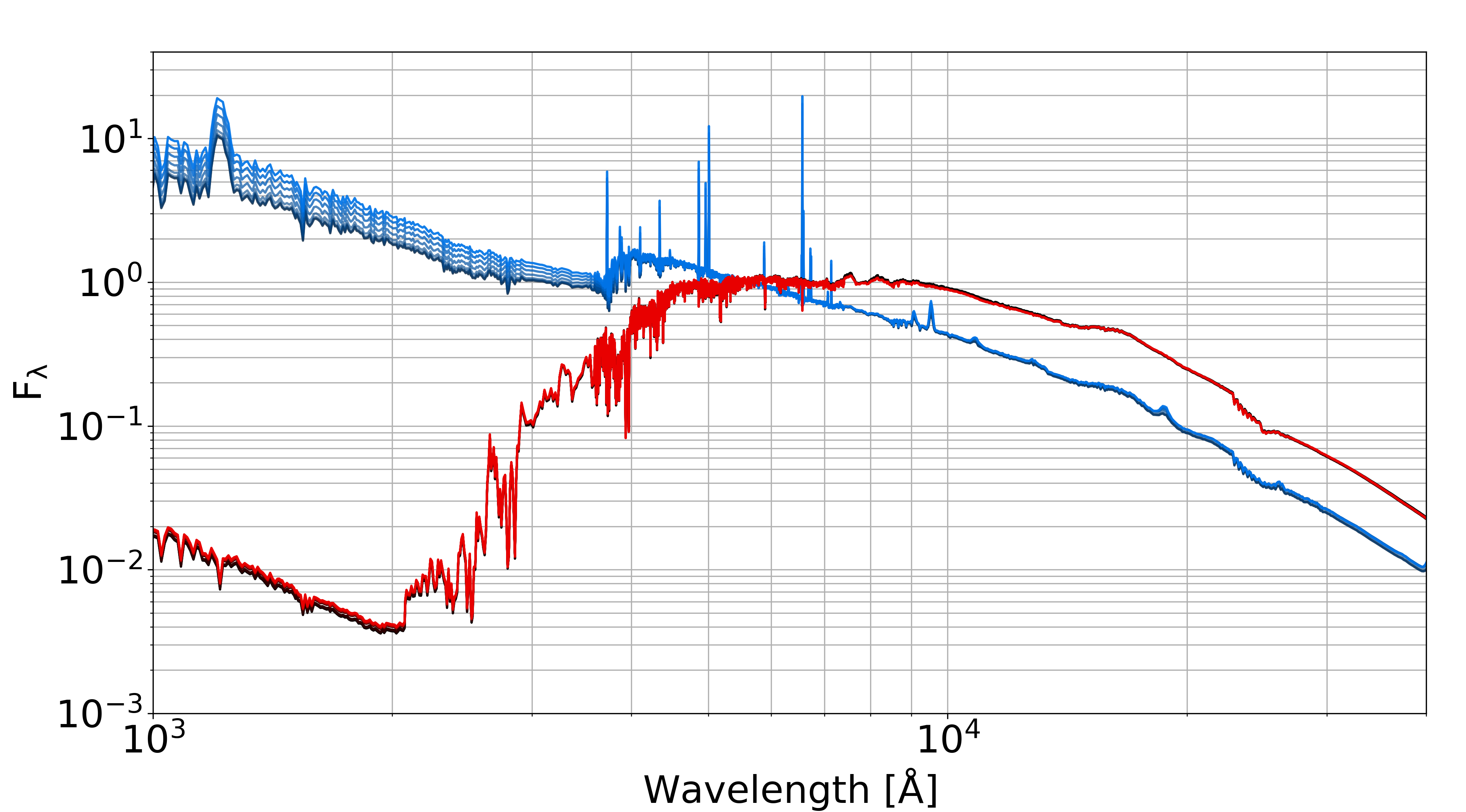}
\caption{FSPS galaxy spectra normalized at 5600 Å for the same galaxy at an age of 1.3 Gyr [blue] and 20 Gyr [red] with increasing temperature [from 20K to 50 K] shown in progressively brighter hues. Increasing temperature results in larger fluxes in the UV [$\lambda <$ 4000 Å] as expected from the additional massive stars. Note the large variability in the young galaxy with temperature while the older galaxy is invariant to temperature-variations. }
\label{fsps_imf}
\end{center}
\end{figure} 

\subsection{Distinguishing a change in the IMF from covariant parameters}
\label{subsec:distinguish}
The greatest impact of increasing $T_{IMF}$ lies in an increased fraction of high-mass stars (Fig. \ref{fsps_imf}), and therefore a bluer spectrum.  Although a bluer spectrum can also be produced by decreasing extinction, extinction is predominantly exponential, while a change in the IMF produces a more complex shape, with different curvature.  The two shapes are therefore distinct, so that a change in IMF allows a better fit. 


Often, the best-fit model using an incorrect shape might be expected to be most incorrect at both extremes.  However, flux at wavelengths longer than $\sim 10^4$\AA \ has only a small contribution from starlight.  As a result, a change in the IMF has negligible effect, since dust and metallicity are fit independent of star formation when using EAZY. Although the true difference is far more significant on the blue end, a best-fit model with a 20K IMF will typically reduce the total $\chi^2$ by varying the extinction (i.e., less dust than is physically present to compensate for a top-heavier IMF). Therefore, the improvement at high $T_{IMF}$ instead lies predominantly at both extremes of the flux most sensitive to starlight.

\begin{figure}[ht]
\begin{center}
\includegraphics[width=\linewidth]{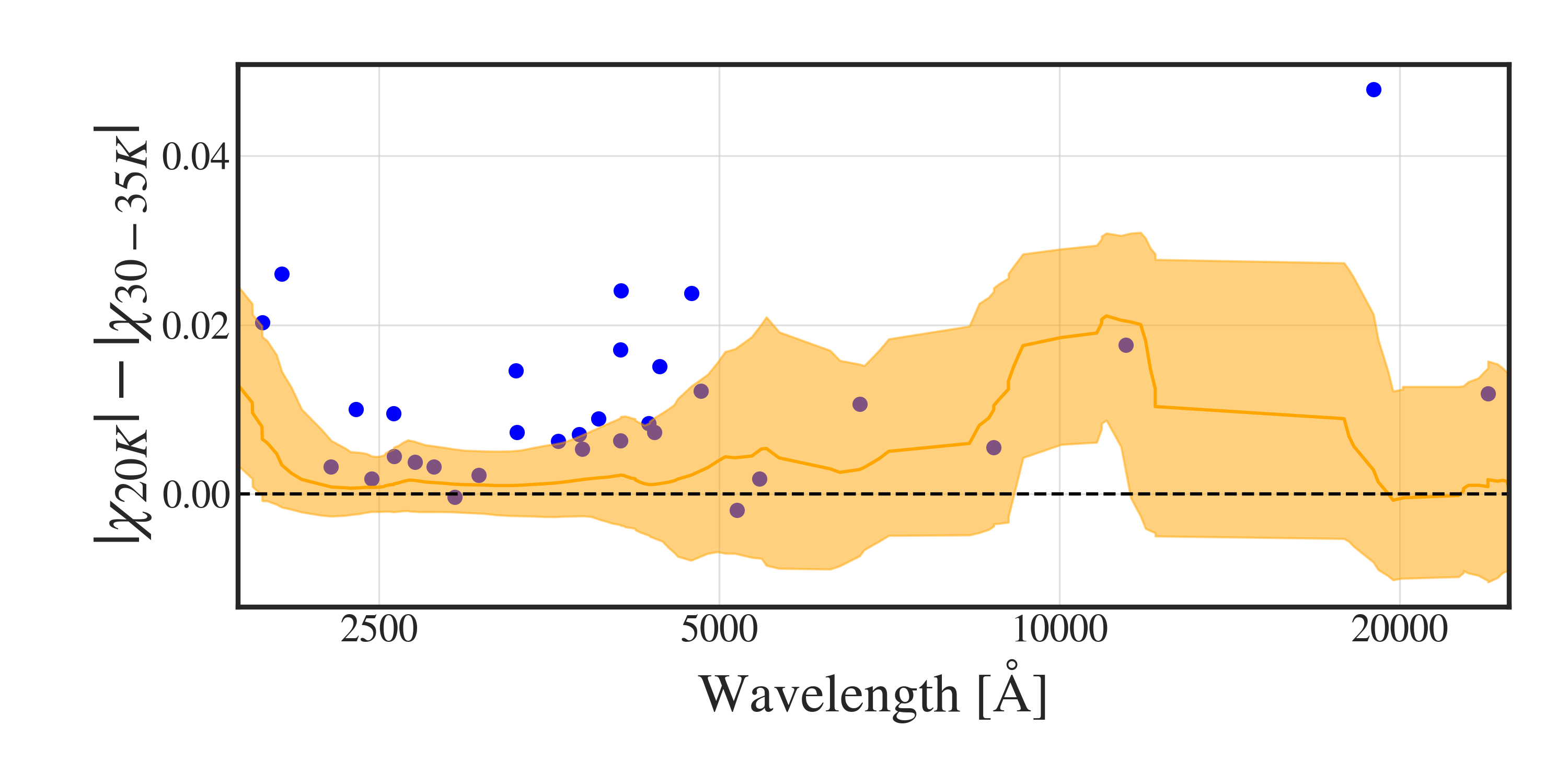}
\caption{Median difference in $\chi$ between the best-fit templates at $T_{IMF}$ and 20K for objects at $1.2 < z < 1.5$ with best-fit $T_{IMF}$ from 30-35K as a function of rest-frame wavelength (orange).  Shading indicates lower and upper quartiles. The difference is shown as a running boxcar with width constant in log-space and set equal to the wavelength. The primary improvement in fit lies in the UV and NIR, likely due to a difference in shape between (power-law) extinction and the more complex effects of a change in $T_{IMF}$.  The fit residuals between a $T_{IMF} = 32$K FSPS template and a best-fit reconstructed spectrum at $T_{IMF} = 20$K are shown for comparison (blue), exhibiting a similar shape} 
\label{fig:residual}
\end{center}
\end{figure}

 In order to confirm this, the residuals of (synthetic) model and observed spectra can be compared. For real spectra, the residuals are calculated by subtracting the best-fit model template (convolved over filter-profiles) from the observed photometric flux and dividing by the reported standard uncertainty in each filter. For model spectra the methodology is identical, with the "observed" photometric flux being a convolution of filter-profiles with a synthetic FSPS model spectrum.  The model FSPS spectrum is that of a star-forming galaxy at $T_{IMF} = 32$K and $z = 1.35$.  The fit was performed using uncertainties in each band that are typical of the overall catalog, even though the model FSPS fluxes were used precisely, with no artificial scatter.  The resulting residuals exhibit a similar shape to the actual residuals for objects at $1.2 < z < 1.5$ with best-fit $T_{IMF}$ from 30-35K (Fig. \ref{fig:residual}). Thus, the improvements in fitting for both model and observed photometry are predominantly at the edges of the starlight-sensitive portions of the SED.  Such a result is consistent with the interpretation that these objects yield superior fits for $T_{IMF} = 30-35$K than for $T_{IMF} = 20$K.

\subsection{Temperature grid and outliers}

Fits are performed over a temperature-grid of IMFs, spaced every 1K from $8\textrm{K}$ to  $50\textrm{K}$. Most objects are well-fit within a significantly tighter interval of temperatures (typically from 20 to 40 K). However, this analysis still presents the larger interval to emphasise the tight and singular relationship between redshift and $T_{IMF}$. The lower limit is 8K, as the steepest portion of the IMF then applies for $M > 0.08 M_{\odot}$. Below approximately $0.08 M_{\odot}$ the objects are no longer stars in the sense that fusion can no longer take place, so to create an even bottom-heavier IMF a different IMF shape would be required (as suggested in \citealt{Bonnell2007}). The upper limit is arbitrarily set at 50 K as no significant population of objects is fit from 45-59 K. With a 60 K upper limit, no significant population would be fit between 50-59 K. At every temperature within the grid, EAZY computes the best-fit linear combinations of templates, a best-fit photometric redshift $z_{phot}$ and a goodness of fit, ${\chi}^2$. The best-fit temperature is the location of the global minimum within the ${\chi}^2$-temperature landscape. 

\begin{figure}[!ht]
\begin{center}
\includegraphics[width=\linewidth]{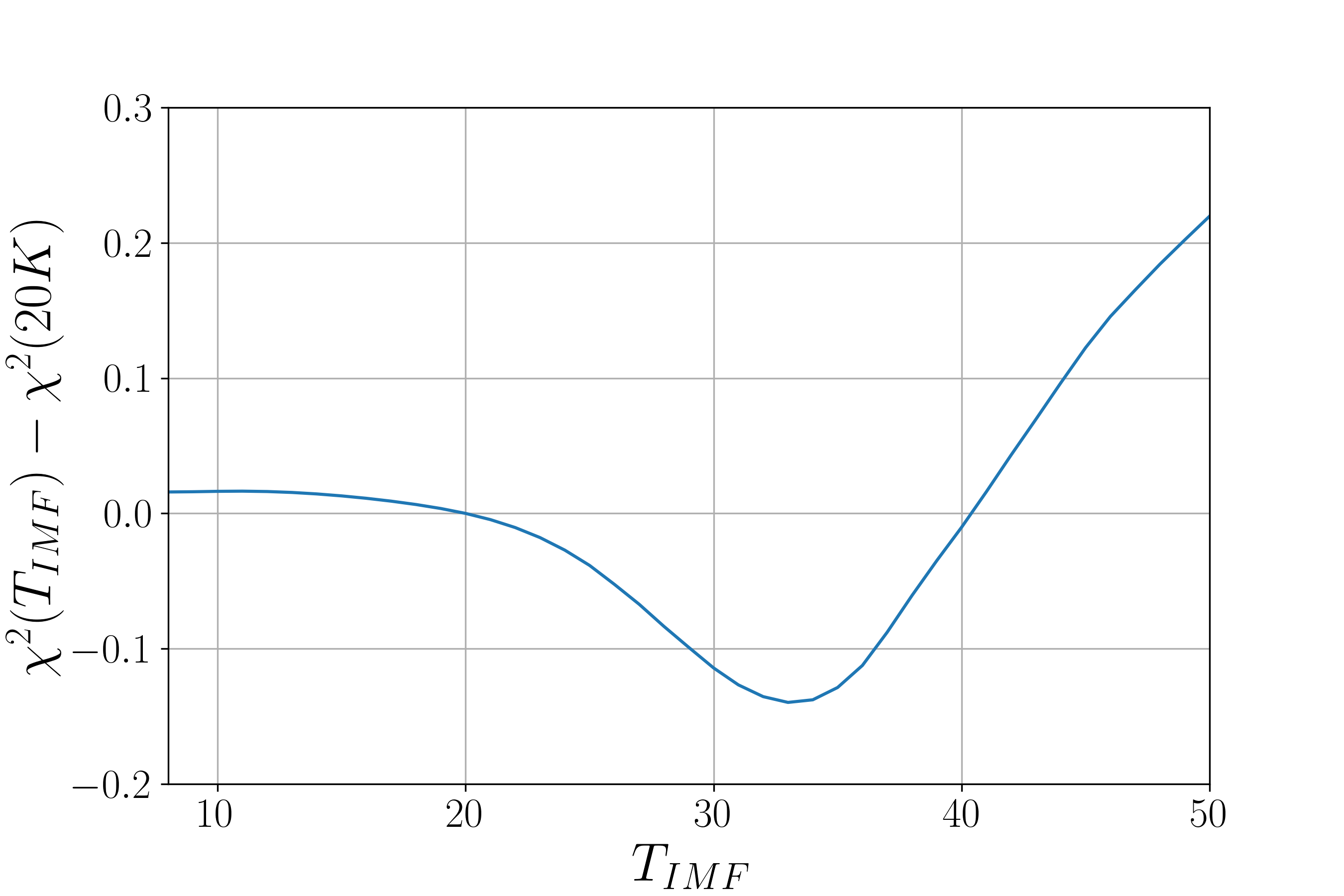}
\caption{An example $\chi^2$-landscape calculated from the best-fit reconstructed SED at different IMF temperatures, with a single clear global minimum. The best-fit temperature does not correspond to a Galactic IMF, but to one significantly top-heavier.}
\label{fig:chi2}
\end{center}
\end{figure}

One concern with this gridded approach is that if there are many local minima and a coarse grid, it is possible that the minimum $\sum \chi^2_n$ may not lie at the global minimum.  This is a known issue more generally in photometric template fitting, and one of the reasons that EAZY chooses to fit using an eigenbasis rather than gridding in all parameters. Fortunately, for most objects in the COSMOS2015 catalog, the $\sum \chi^2_n$ surface typically indicates only one local minimum, which is also the global minimum (e.g., Fig. \ref{fig:chi2}).  

\begin{figure}[!ht]
\begin{center}
\includegraphics[width=\linewidth]{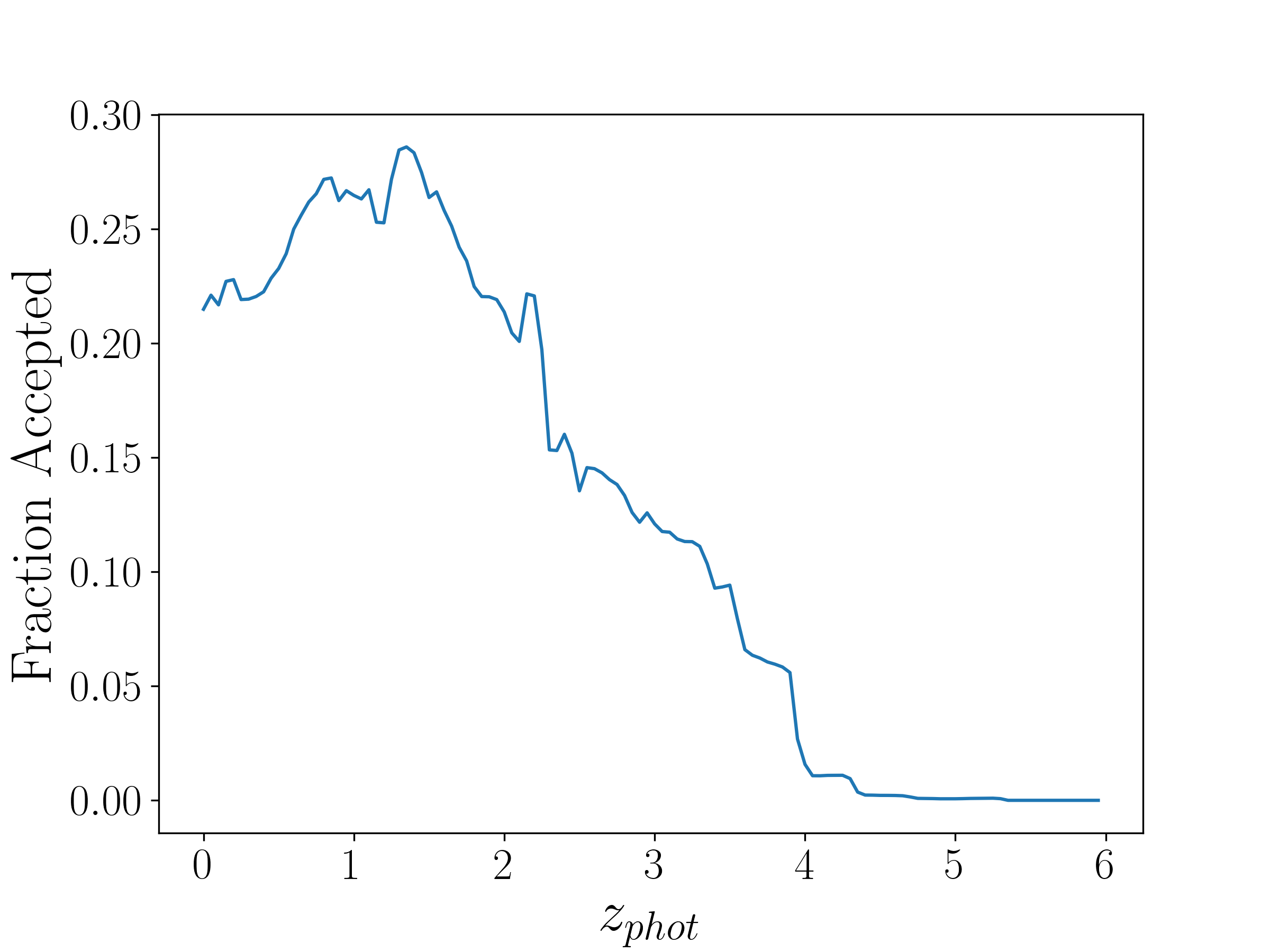}
\caption{The fraction of objects in COSMOS2015 accepted under the SNR cut of 10 in V-band and with a minima in the temperature range investigated relative to all sources as a function of photometric redshift.}
\label{fig:qualitycut}
\end{center}
\end{figure}

For 60\% of all objects in the catalog there is a single local minimum within the $\chi^2$-temperature landscape, which is also the global minimum. For 40\% of the catalog, there are several local minima or the only minimum lies at bounds of the temperature range (at 8K or 50K).  A closer examination of these cases reveal that they typically have poorly-constrained photometry with a low SNR. With the SNR-cut instituted above around 80\% have a single local minimum within the $\chi^2$-temperature landscape\footnote{There is also a small population of very high-mass star-forming galaxies for which there may be a physical explanation for multiple minima, as described in Paper III.}. Furthermore, the number of objects fit at the temperature bounds is tightly correlated with the number of photometric bands used for template-fitting. Decreasing the number of photometric bands used yields a higher fraction of outliers. Conversely, objects observed in a large number of bands are often well-constrained. Objects with poorly-constrained photometry yield poorly constrained $T_{IMF}$, so for the remainder of this work objects with no clear local minimum are cut from the sample as low quality.  As might be expected, this quality cut affects a greater number of objects at lower redshift but a greater fraction of objects at higher redshift (Fig. \ref{fig:qualitycut}).  

A possible concern is that cutting outliers in $T_{IMF}$ may bias towards younger galaxies.  Constraints on the IMF are stronger for younger stellar populations than for older ones, because changes in the IMF lead to larger changes in the resulting spectral energy distributions (SEDs) for younger galaxies (Fig. \ref{fsps_imf}).  As might be expected, in both cases, increasing temperature leads to more massive stars and thus more emission in the UV, while emission in the optical and IR remains unchanged.  However, the older galaxy displays less variability with temperature because massive stars have shorter lifetimes, so most of these massive stars are no longer present.

\section{IMF Temperatures for the COSMOS2015 Catalog}
\label{sec:results}
Using the techniques described in the previous section, a modified version of EAZY was run to find a best-fit IMF and other galactic parameters for every object in the COSMOS2015 catalog with sufficient signal to noise in the V-band. As described in \S~\ref{subsec:catalog}, because constraining the IMF involves fitting for an additional parameter, doing so successfully requires a stronger quality cut than in previous COSMOS2015 analysis.  Of the 518.403 galaxies objects for which \citet{Laigle2016} were able to constrain the redshift, only 139.535 had sufficient signal to noise to constrain the IMF.  This cut is evaluated in more detail in \ref{subsec:qualitycuts} in order to confirm that it does not introduce an additional bias.  

\begin{figure*}
  \includegraphics[width=.62\linewidth]{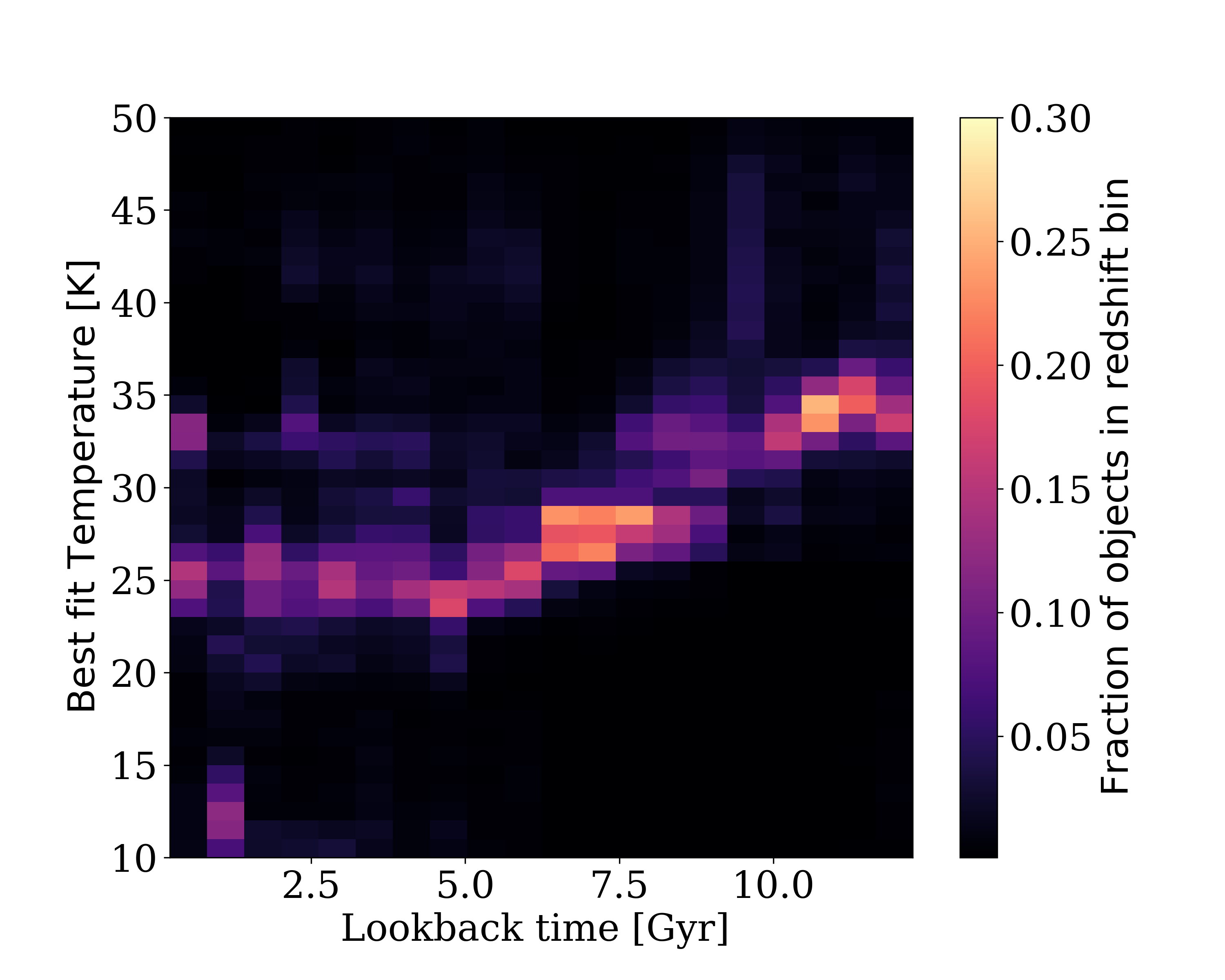} \quad
  \includegraphics[width=.35\linewidth]{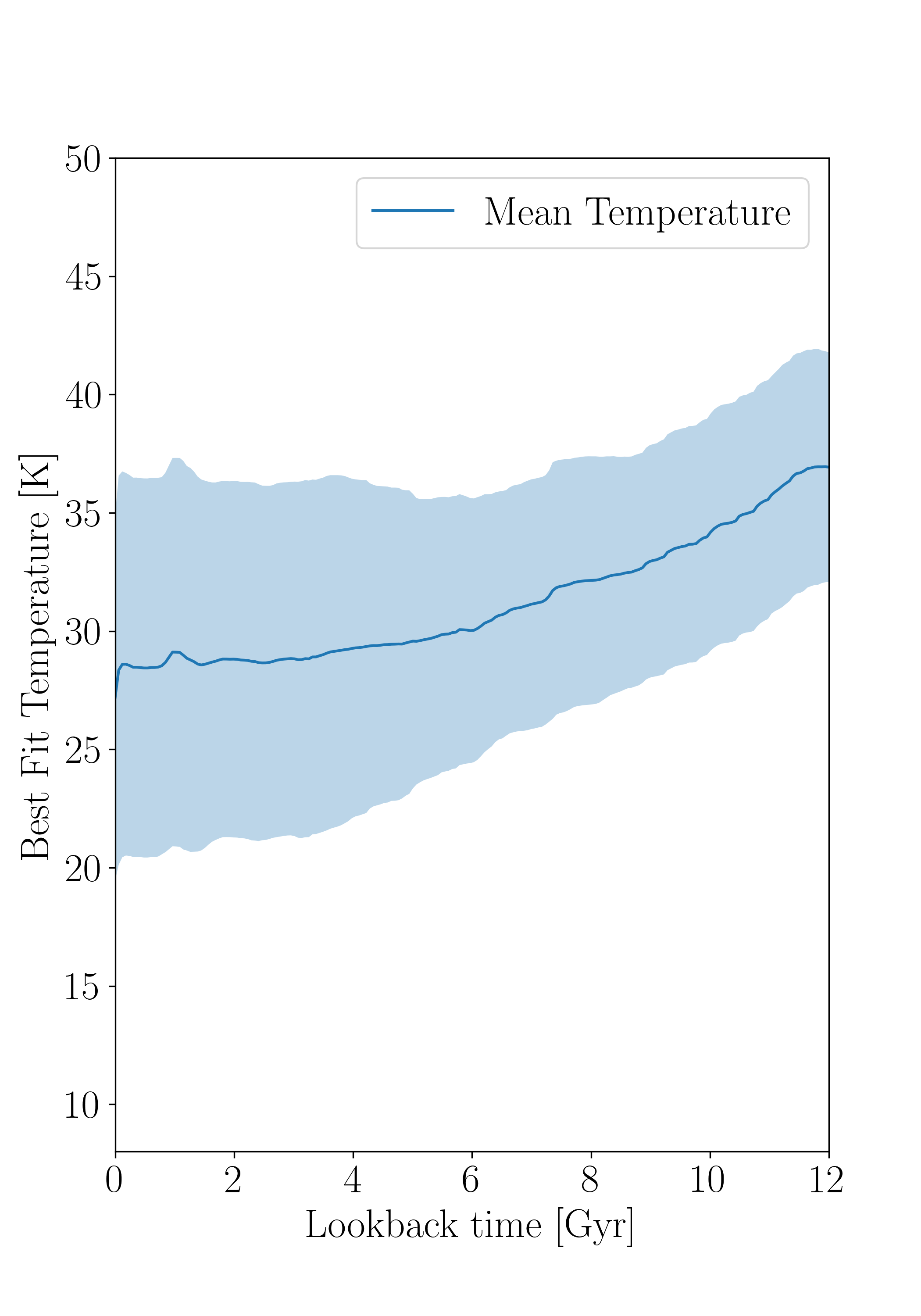}
\caption{Left: Best fit temperature from 10-50 K versus lookback-time from a sample of 139,535 
COSMOS2015 galaxies with SNR$>$10 in the V-band \citep{Laigle2016}. At each redshift, the distribution is individually normalised, in order to emphasise the temperature distribution at all redshifts. With increased redshift fewer galaxies are fit at lower temperatures. Right: Boxcar smoothed mean with standard deviation of best-fit gas temperature at different lookback-times [with mean determined from objects in 2 Gyr width age bins and not including galaxies fit at the bounds of temperature range]. The mean temperature increases from \ $\sim 28$ K to \ $\sim 36$ K from present to 12 Gyr, while the spread decreases.} 
\label{fig:temp-z-all}
\end{figure*}

\subsection{A Redshift-IMF Relation}
\label{subsec:singleimf}

Perhaps the most striking result is that nearly every galaxy in the COSMOS2015 sample is best fit not using a Galactic IMF, but using one at $T_{IMF} > 20$ K, or top-heavier than the Milky Way (Fig. \ref{fig:temp-z-all}). 

Further, there is a strong correlation between redshift and the best-fit $T_{IMF}$, with the characteristic $T_{IMF}$ rising from $\sim 25$ K at $z \sim 0$ to $\sim 35$ K by $z \sim 4$.  If $T_{IMF}$ is indeed an indicator of conditions in star-forming molecular clouds, it appears that these clouds have similar conditions across most galaxies at the same redshift.  Further, those conditions change with redshift, implying that there is a cosmic synchronization to key aspects of galactic evolution.  

These results contradict the current assumption that the IMF is universal and constant, so that a Galactic IMF can be used to analyze all galaxies at all redshifts.  Thus, one significant implication is that photometric observations of most galaxies may need to be reinterpreted.  This might change key inferred properties, including not just stellar masses and star formation rates but also extinction, age, metallicity, etc., which may have been incorrectly inferred for most galaxies.  This is discussed in more detail in \S~\ref{subsec:inferredquantities}.  The specific implications for the star-forming main sequence are discussed further in Paper II, and for quiescent galaxies and quenching mechanisms in Paper III.  

Finally, although gas temperatures have historically not been possible to observe for most galaxies, dust-temperatures present an independent probe of the trend seen in Fig. \ref{fig:temp-z-all} and have been estimated for a broader ensemble of evolving galaxies. The inferred characteristic dust temperature of galaxies on the star-forming main sequence decrease from approximately 35 K to 25 K from $z = 2$ toward the present time \citep{Magnelli2014,Magdis2017}.  If the temperatures of dust and cold gas clouds are similar, then one would expect a similar trend in gas temperature, which is consistent with the observed evolution in $T_{IMF}$.  It should be noted that the two cannot be compared directly, since the luminosity averaged dust temperature is dominated by hotter dust, while the gas temperature is mass-averaged as it is inferred from the IMF.  However, despite the discrepancies, $T_{IMF}$ measurements are entirely independent of dust temperature estimates, so it is striking that the two produce consistent results.

\subsection{Physical Populations}
\label{subsec:subpopulations}

Galaxies are known to fall into subpopulations with distinct properties.  Perhaps the strongest separation is between star-forming and quiescent galaxies, a classification which can easily be determined from color for most galaxies \citep{Williams2009,Arnouts2013}.  Cool gas in star-forming galaxies is used to produce stars.  Various mechanisms have been proposed for quiescence, either depleting the cool gas or producing conditions under which cool gas exists but cannot efficiently produce stars.  It is therefore natural to wonder whether both star-forming and quiescent galaxies exhibit similar IMFs.

\begin{figure}
  \includegraphics[width=\linewidth]{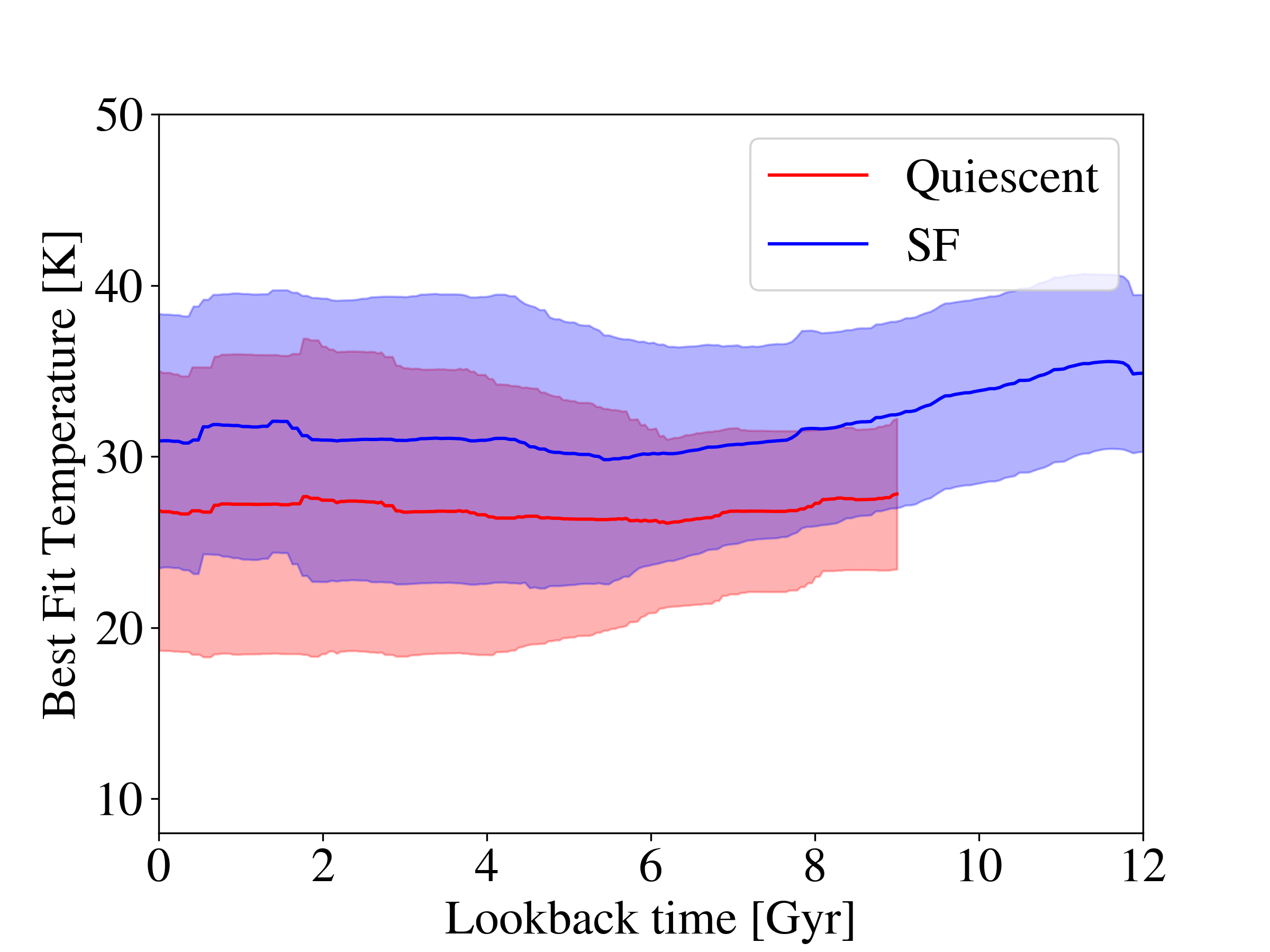} \quad
  \caption{ Box-car smoothed mean and spread of inferred temperature for galaxies in age bins of width 2 Gyr. Quiescent galaxies display Milky Way like IMFs and lie on the lower range of temperatures spanned by star-forming galaxies. Well-measured quiescent galaxies are sparse beyond 9 Gyr, with significant numbers of star-forming galaxies observed up til 12 Gyr.}
\label{fig:star-forming}
\end{figure}

Using UVJ selection, the galaxies in the COSMOS2015 catalog were divided into star-forming and quiescent populations.  The distribution of best-fit $T_{IMF}$ for the two populations differs (see Fig. \ref{fig:star-forming}).  The continuous range of temperatures at a given redshift seen in Fig. \ref{fig:temp-z-all} consists of a combination of cooler quiescent galaxies and a range of slightly hotter star forming galaxies.  Thus, a selection based on $T_{IMF}$ provides a surprisingly effective separation of the two classes.  The coolest galaxy at any redshift is typically quiescent. 

An examination of best-fit temperatures for synthetic spectra (see \S~\ref{subsec:qualitycuts}) implies that the total uncertainty on individual temperature measurements is approximately 10K. Thus, many individual objects cannot be selected as quiescent or star-forming from their best-fit $T_{IMF}$ alone.  However, the uncertainty on the mean of a population of galaxies is naturally far better constrained given the Central Limit Theorem. For $0<z<1$, there are 7.800 quiescent galaxies and 68.000 star forming galaxies, so the star-forming galaxies mean $T_{IMF}$ is more than 30$\sigma$ higher than the quiescent sample. However, while the statistical uncertainty on an ensemble is small, it is difficult to conclusively show that this separation truly comes from a difference in temperature.  Because the EAZY template basis only spans a limited subspace of possible galaxy spectra, an alternative explanation could be that typical, e.g., quiescent spectra lie outside that subspace, while typical star-forming spectra are better modeled by that limited basis.  However, these results are consistent with the naive prediction that a less active galaxy should be colder due to less stellar light and lower cosmic rays production. This is also consistent with measured dust temperatures, which tend to increase towards higher SFR at fixed stellar mass and redshift \cite{Magnelli2014}. 

\begin{figure}[!ht]
  \includegraphics[width=\linewidth]{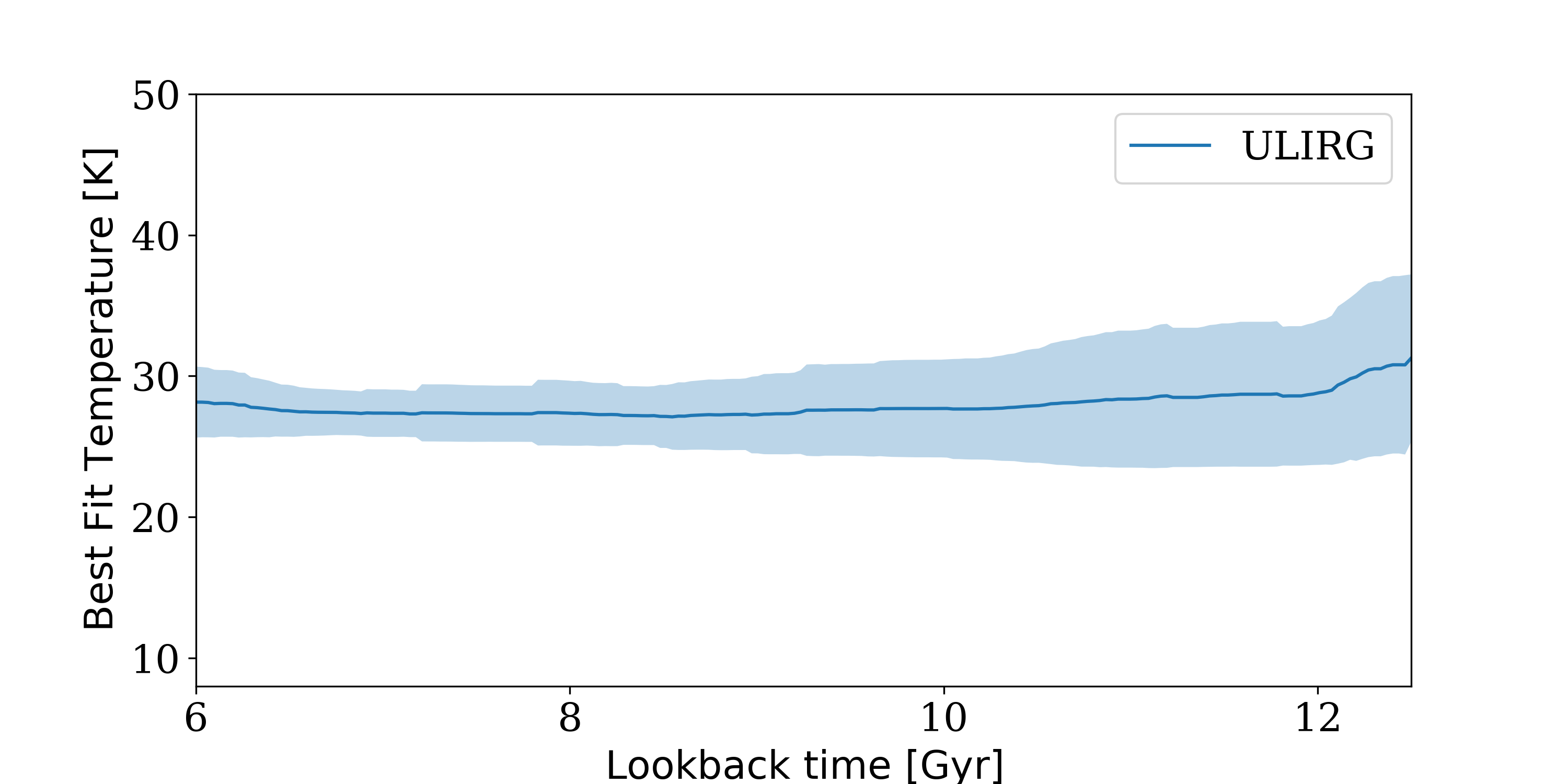} \quad
  \caption{ Box-car smoothed mean and spread of inferred temperature for ULIRGs with age bins of width 2 Gyr. The average temperature is consistent with being independent for lookback-times ranging from 6 to 12.5 Gyr. }
\label{fig:ulirgs}
\end{figure}

A smaller subpopulation are the the ultra-luminous infrared galaxies (ULIRGS), hypothesized to be caused by mergers of gas-rich galaxies \citep{Lonsdale2006}.  The typical best-fit IMF of ULIRGs (Fig. \ref{fig:ulirgs}) is nearly independent of redshift within the range of lookback times investigated from 6 to 12.5 Gyr.  In that respect, ULIRGs behave differently than typical star-forming galaxies, which have hotter gas towards high redshift.  This might reinforce the belief that ULIRGs are governed by a distinct process from those which drive the star-forming main sequence, and thus typical star-forming galaxies.  This redshift-independence of temperature has also been measured in dust-temperatures of ULIRGs \citep{Bethermin2015}, again suggesting that gas and dust temperatures are correlated, either because they are in equilibrium or due to feedback processes.

\subsection{Comparing Temperature Evolution with Theory}
\label{subsec:crpredictions}

Although previous template fitting techniques have generally assumed a Universal IMF, there have been several solid theoretical reasons to expect that this might not be the case.  It is therefore natural to consider whether the best-fit $T_{IMF}$ are consistent with what might have been expected theoretically.

Within the Milky Way, the typical temperature of cold gas in molecular clouds is set primarily by stellar radiation.  Different molecular clouds may have different temperatures, depending upon their proximity to luminous stars.  The most common case will be temperatures set by low-mass stars, as the Galactic SFR is relatively low, so the probability of a high-mass star in close proximity to a molecular cloud is small.  
However at higher SFR, as was the case for much of the Milky Way's history, the dominant contribution may instead come from cosmic rays, which permeate the entire galaxy and therefore can heat clouds uniformly \citep{Papadopoulos2010a,Papadopoulos2010b}.  At $z > 6$, beyond the redshift domain discussed in this work, the CMB temperature also exceeds that of typical Galactic molecular clouds \citep{Jermyn2018}.  In both cases, clouds would have similar temperatures throughout the galaxy.  In this work, a single characteristic temperature is fit due to the limited information available, so that the resulting best-fit parameters follow the standard convention of treating galaxies as monolithic, with one SFR, $T_{IMF}$, extinction curve, metallicity, etc.

In the \citet{Steinhardt2020} model of galaxy evolution the feedback processes between cosmic ray heating from the deaths of massive stars and the temperature dependence of the IMF result in an attractor solution with a common $T_{IMF}$. They predict that $T_{IMF}$ should be nearly constant for star-forming galaxies at fixed redshift, and declining towards lower redshift.  Furthermore, quiescent galaxies should have lower $T_{IMF}$ than star-forming galaxies, but should not continue to drop past quiescence, as the dominant heating source at low SFR is instead starlight from low-mass stars. These predictions are consistent, at least qualitatively, with the results found here, but there are sufficient free parameters that \citet{Steinhardt2020} were unable to make robust quantitative predictions with their model.

There are also several other observations which hint at the same conclusion.  Although absolute calibration is difficult, so that inferred dust temperatures vary quantitatively between models, a corresponding attractor dust temperature and decrease in dust temperatures toward low redshift is seen on the star-forming main sequence \citep{Casey2012,Magnelli2014,Bethermin2015,Magdis2017,Cortzen2020}.  This, too, would suggest that that the IMF typically becomes more top-heavy at larger redshifts.

Notably, this is one of many other tight correlations known in galaxy evolution including the star-forming main sequence, a tight correlation between star formation rates and the existing stellar mass at fixed redshift (cf. \citet{Peng2010,Speagle2014,Tomczak2016}).  Such a correlation might imply a universality to star formation with a common evolutionary track \citep{Steinhardt2014b}, but alternatively could be explained by averaging over a large number of shorter bursts \citep{Kelson2014}.  The consistency of $T_{IMF}$ is clearly consistent with the former explanation.  For the latter, it would appear to be inconsistent with bursts on a galaxy-wide scale, for which $T_{IMF}$ would vary significantly.  However, a galaxy composed of several subpopulations, which are individually either starbursting or not at any given time, could still produce a nearly-constant $T_{IMF}$ when the entire galaxy is fit simultaneously and treated as monolithic.  Testing this hypothesis will require fitting resolved subpopulations of nearby star-forming galaxies in order to determine whether they all have the same $T_{IMF}$.

\subsection{Reinterpreting Galactic Properties}
\label{subsec:inferredquantities}

\begin{figure}
  \includegraphics[width=.95\linewidth]{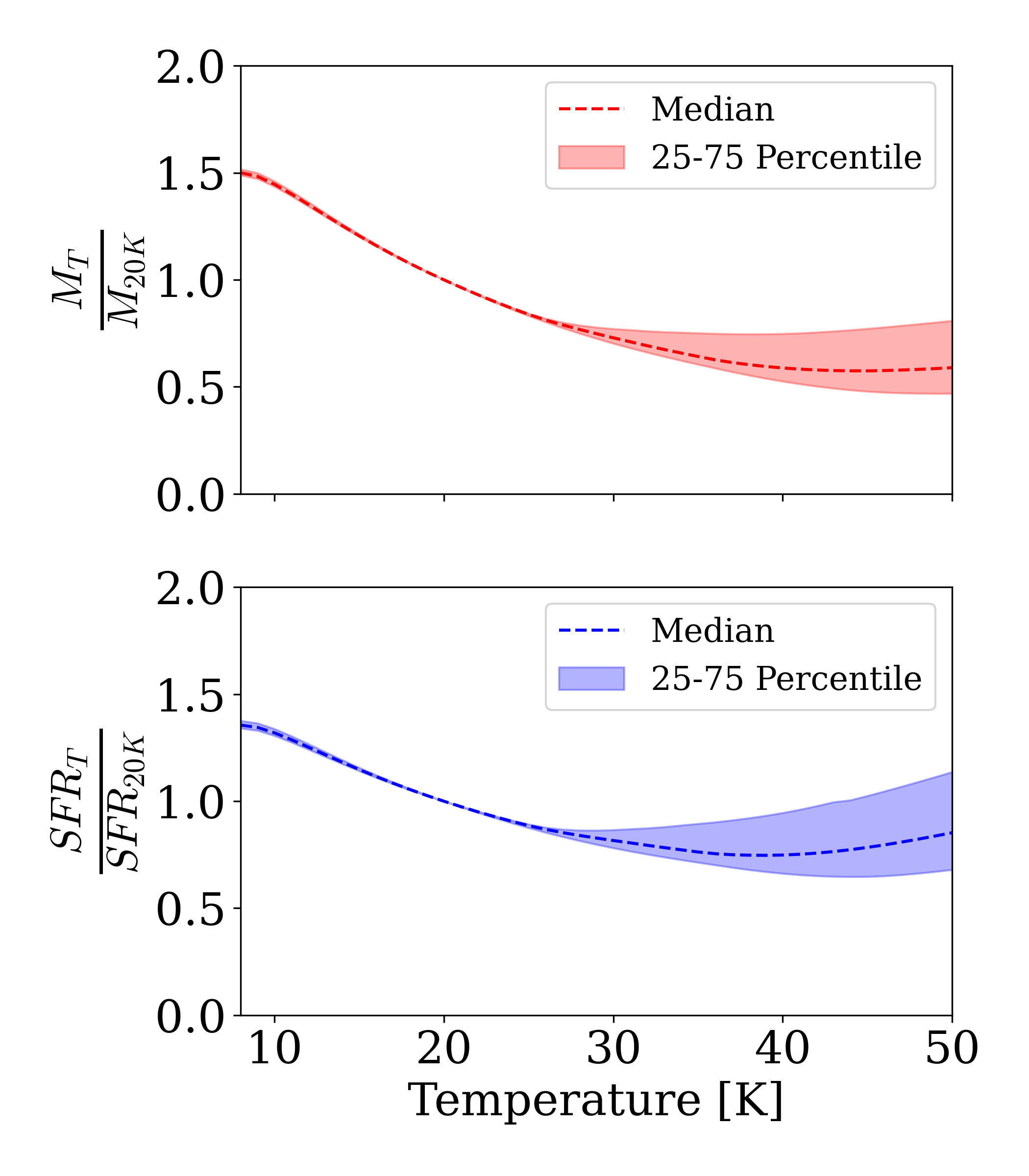}
\caption{Best fit stellar mass (M) and star formation rate (SFR) from EAZY template-fitting for individual galaxies as a function of $T_{IMF}$. The dashed line is the median with shading indicating first and third quartiles. Any observed galaxy will only have a single best-fit IMF, but the stellar mass and SFR may be biased by an erroneous choice of IMF. Stellar Mass decrease monotonically with increasing temperature as we would expect from a top-heavier IMF, while SFR increases slightly beyond 40 K.} 

  \label{fig:gal_param}
\end{figure}

Nearly every galactic property inferred from photometric template fitting is sensitive to the low-mass stellar population which is not luminous enough to be seen, and therefore depends upon the assumed IMF.  Thus, changing the IMF will also change these inferred properties.  Following the standard EAZY parameter reconstruction  \citep{Brammer2008}, the galactic properties of individual templates can be propagated by a luminosity-weighted sum of the basis templates into the properties of any photometric observations.  A first approximation of this effect can be seen in Fig. \ref{fig:gal_param}, where the best-fit parameters are shown for identical sets of photometric colors for different $T_{IMF}$. 

Perhaps the most straightforward changes are those in stellar mass and IMF for small variations in the Galactic IMF ($T_{IMF} \sim 20$ K).  Bottom-heavier IMFs (lower $T_{IMF}$) produce larger inferred $M_*$ and SFR, as more stellar mass is 'hidden' in small stars.  The luminosity and mass are respectively dominated by the upper and lower end of the stellar mass distribution, so a bottom-heavy IMF decreases the luminosity relative to the mass. Thus, for the same measured luminosity, a greater mass is inferred. Correspondingly, a top-heavier IMF will have lower stellar masses and SFR than at $T_{IMF} = 20$ K. Given the evolution evident in Fig. \ref{fig:temp-z-all}, this implies that previous analyses may have systematically overestimated stellar masses, particularly towards high redshift, with the true stellar masses being around 40 \% less than predicted for a typical $z=2$ galaxy.  Naturally, the exact bias in inferred properties will depend on the parameterization of a variable IMF.

Although at $T_{IMF} < 30$ K, the trends for stellar mass and SFR as a function of $T_{IMF}$ are qualitatively similar, above 30 K they diverge.  The SFR is no longer monotonic, and overall exhibits a weaker temperature dependence than $M_*$.  Notably, this means that varying the IMF will change the specific star formation rate (sSFR).  The implications of this for the star-forming main sequence are described in detail in Paper II.  

The counterinuitive increase in best-fit SFR towards higher $T_{IMF}$ is due to covariances with other galactic properties which are fit simultaneously.  A higher $T_{IMF}$ produces more high-mass stars, and thus a bluer stellar population.  Since the spectral tilt of the underlying photometry has not changed, instead extinction is increased to compensate, and the best-fit combination becomes both higher-SFR and higher-extinction than before.  Of course, these covariances also present a significant concern.  If these properties were fully degenerate, then it would never be possible to determine the true IMF, SFR, or extinction.  Fortunately, they are only partly degnerate, as shown in \S~\ref{subsec:degeneracies}.  

Finally, note that although there is a strong dependence between $T_{IMF}$ and other parameters, there is always a single, best-fit combination of $T_{IMF}$, $M_*$, SFR, etc.  The dependence shown in Fig. \ref{fig:gal_param} is useful to estimate the extent to which properties calculated assuming a Galactic IMF might be incorrect.  For many high-redshift galaxies, a Galactic IMF can be rejected, although for some galaxies, particularly a lower redshift, it cannot be.

\section{Robustness of IMF Measurements}
\label{sec:tests}
Photometric template fitting uses limited information to fit many parameters with high covariance.  As a result, quantities inferred for individual galaxies are rarely tightly constrained.  The introduction of an additional parameter, $T_{IMF}$, only adds to the difficulty and weakens the constraints.  A stronger quality cut is needed to select objects for which $T_{IMF}$ can be measured than in the previous COSMOS2015 catalog, with fewer parameters.  In this section, the robustness of these measurements is evaluated in more detail.

\subsection{Significance and Uncertainty Estimation}
\label{sec:sig}
The $\chi^2$ landscapes produced by photometric template fitting can be remarkably complex, characterised by a large number of local minima, which are often wide but shallow \citep{Speagle2016}.  In addition, because the set of physical models spanned by templates is only expected to approximate the true spectrum, rather than contain it, the errors will not be chi-squared distributed, and therefore the likelihood distribution for $T_{IMF}$ cannot be inferred using a chi-squared test.  

Instead, here uncertainties are estimated using a Monte Carlo simulation.  Each band has a well-quantified uncertainty on the flux within that filter.  Drawing repeatedly from that distribution and finding best-fit parameters produces a resulting distribution of $T_{IMF}$ indicates a typical $T_{IMF}$ uncertainty of 4 K.  The standard deviation in $T_{IMF}$ observed at fixed redshift (see Fig. \ref{fig:temp-z-all}) ranges from 8 K (at $z\approx 0$) to 5 K (at $z \approx 3$).  Importantly, the uncertainty in $T_{IMF}$ for individual galaxies is smaller than the spread around $z \sim 0$ (see Fig. \ref{fig:temp-z-all}), suggesting that galaxies in the local Universe are not consistent with a single $T_{IMF}$.  However, at high redshift the spread in temperature is consistent with being derived from a population at a single temperature.  One possible explanation, described in Paper II, is that star-forming and quiescent galaxies typically lie at different $T_{IMF}$.  If so, a high-redshift population composed nearly exclusively of star-forming galaxies would have a single temperature but one in the local Universe, including significant fractions of both star-forming and quiescent galaxies, would not.




\subsection{Degeneracies and Covariances}
\label{subsec:degeneracies}
A significant concern in constraining the IMF is the possibility of degeneracies in the spectrum with other galactic properties.  In particular, there is in principle a complete denegeracy between the IMF and the star formation history (SFH).  An identical observed stellar population can be produced from different IMFs by varying the star formation history.  This idea has also been used, assuming a Galactic IMF, to determine extragalactic SFHs \citep{Panter2003,Sanchez2019}.  The most recent SFR is determined from the highest-mass objects, then that population subtracted off and the next-most-recent SFR is determined from the highest-mass residual, and so on.  However, such a solution is possible for a wide range of IMFs.  In principle, a particularly incorrect IMF may require a negative SFR at some times, but in practice, many reasonable choices of IMF produce non-negative results.

\begin{figure}[ht]
\begin{center}
\includegraphics[width=\linewidth]{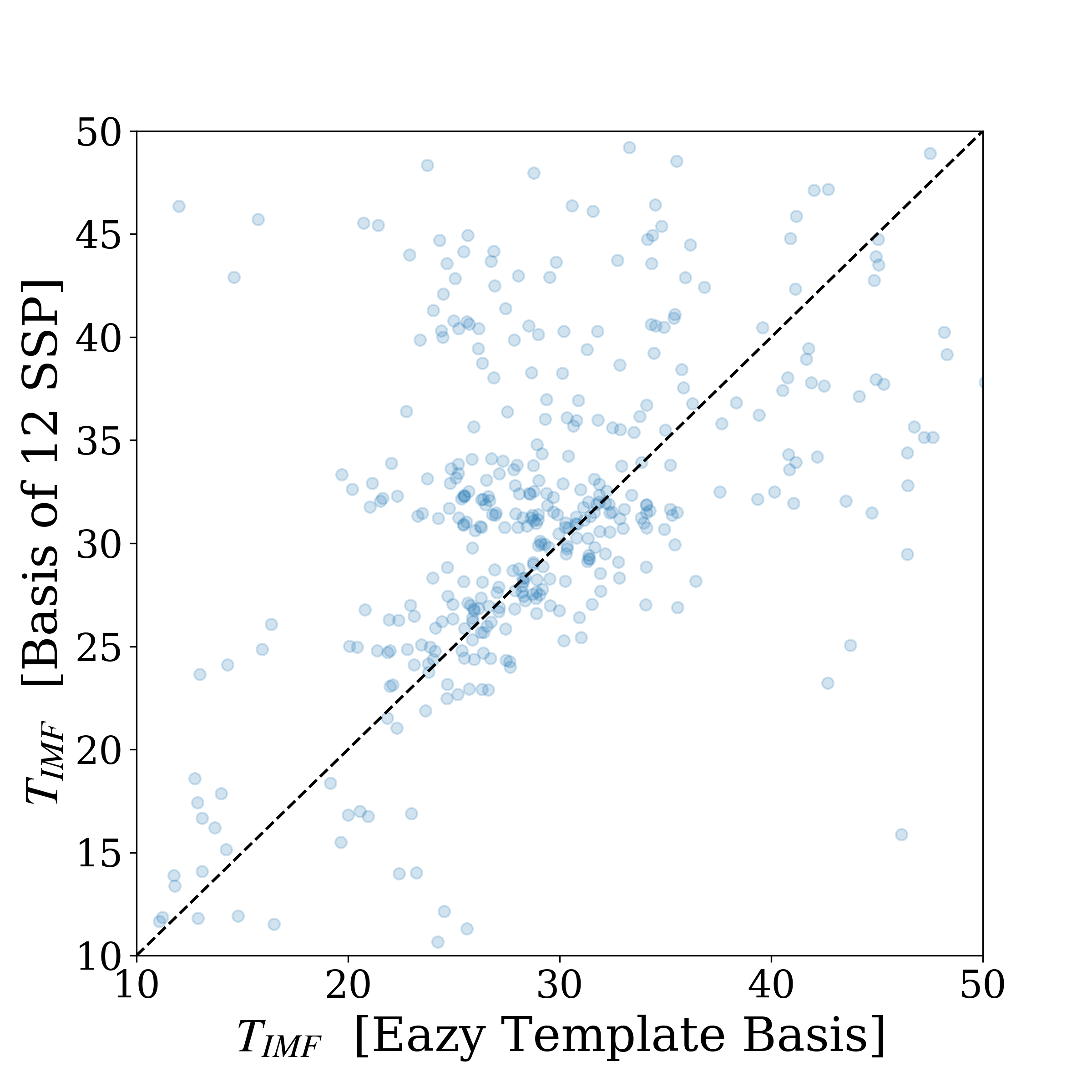}
\caption{Inferred $T_{IMF}$ for 1,000 objects given a basis of 12 representative SSP templates and the EAZY template-basis of composite delayed $\tau$-model SFH history. Regardless of assumption in star formation history similar temperatures are inferred with a correlation coefficient of 0.5 for objects found with a $\chi^2$-minima within the temperature range. } 
\label{fig:SFH-comp}
\end{center}
\end{figure}

Here, this degeneracy is broken by assuming a shape for the SFH.  In the standard EAZY methodology, templates are generated using a delayed-tau SFH, with a range of different ages for the stellar population \citep{Brammer2008}.  For this single shape of the SFH, it is then possible to uniquely determine a best-fit IMF.  Changing the assumed shape of the SFH will, of course, change the resulting best-fit parameters.  As an extreme example, one can generate templates using a history consisting only of a single starburst, producing a single stellar population.  However, applying these SSP-templates to observed photometry yields a best-fit $T_{IMF}$ similar to those with the delayed-tau model (see Fig. \ref{fig:SFH-comp}). Thus, the inferred IMF remains relatively robust regardless of varying assumptions of SFH. 

\begin{figure*}[ht]
\begin{center}
\includegraphics[width=\linewidth]{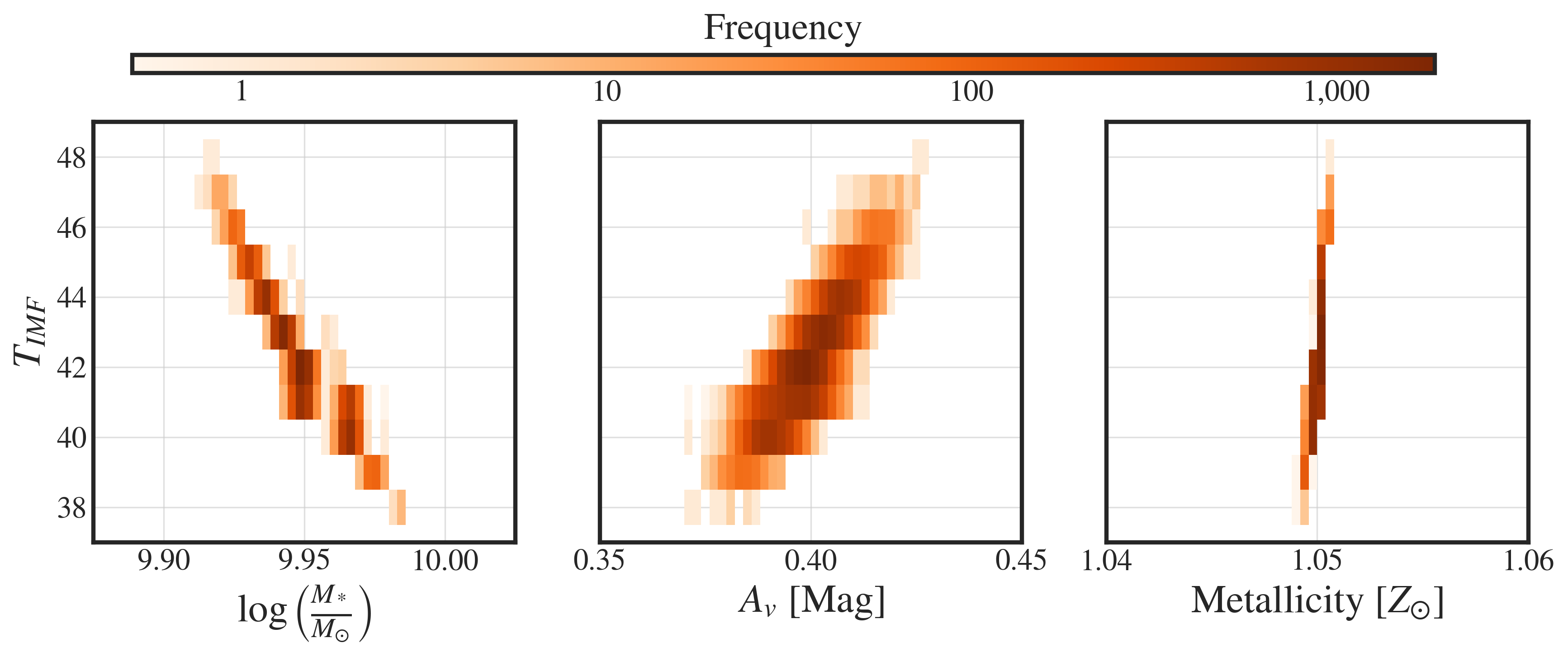}
\caption{ Frequency of best fit $T_{IMF}$ versus metallicity, extinction coefficient and stellar mass for a single object in COSMOS2015 with 10,000 unique perturbations of observed photometric flux. The perturbed fluxes are generated by sampling from a normal distribution with mean observed flux and standard deviation flux error, as reported in the catalog. $T_{IMF}$ displays respectively negative and positive covariance with the inferred mass and dust extinction, with no clear correlation to metallicity. } 
\label{fig:degeneracies}
\end{center}
\end{figure*}

An additional concern are the covariances between $T_{IMF}$ and properties such as metallicity and dust. For example, to first approximation, a higher $T_{IMF}$ produces a bluer spectrum for the same physical properties, much as lower extinction would.  
To investigate this, the same Monte Carlo simulation described in the previous section was used, comparing the distributions of pairs of physical properties (see Fig. \ref{fig:degeneracies}).  As expected, there is large degeneracy between $T_{IMF}$ and $M_*$, as the IMF is a near-direct scaling between the bright end of the stellar mass function and the remainder.  Furthermore, an increasingly top-heavier IMF is offset by increasing extinction. Indeed, this trade-off between extinction and IMF could produce the characteristic shape of residuals discussed in \S \ \ref{subsec:distinguish}. Finally, there are no significant degeneracies between $T_{IMF}$ and metallicity. However, metallicity is derived from the strength of lines and is poorly constrained from photometry. Therefore, the small statistical variation is not representative of the systematic uncertainties associated with constraining $Z$ for template-fitting \citep{Mitchell2013}.

Additionally, the co-variance between $T_{IMF}$ and the inferred $z_{phot}$ can be estimated by examining the 23.000 objects with additional spectroscopic redshift in COSMOS2015 catalog \citep{COSMOS2015}. Here, the scatter between spectroscopic redshift and photometric redshift, $z_{phot}-z_{spec}$, is unchanged with a variable $T_{IMF}$. Thus, the improvements in the fit does not decrease the accuracy of the photometric redshift. Physically, the photometric redshift in template fitting is typically dominated by breaks in the spectrum, while $T_{IMF}$ is primarily fit from the continuum. 

\begin{figure}[ht]
\begin{center}
\includegraphics[width=\linewidth]{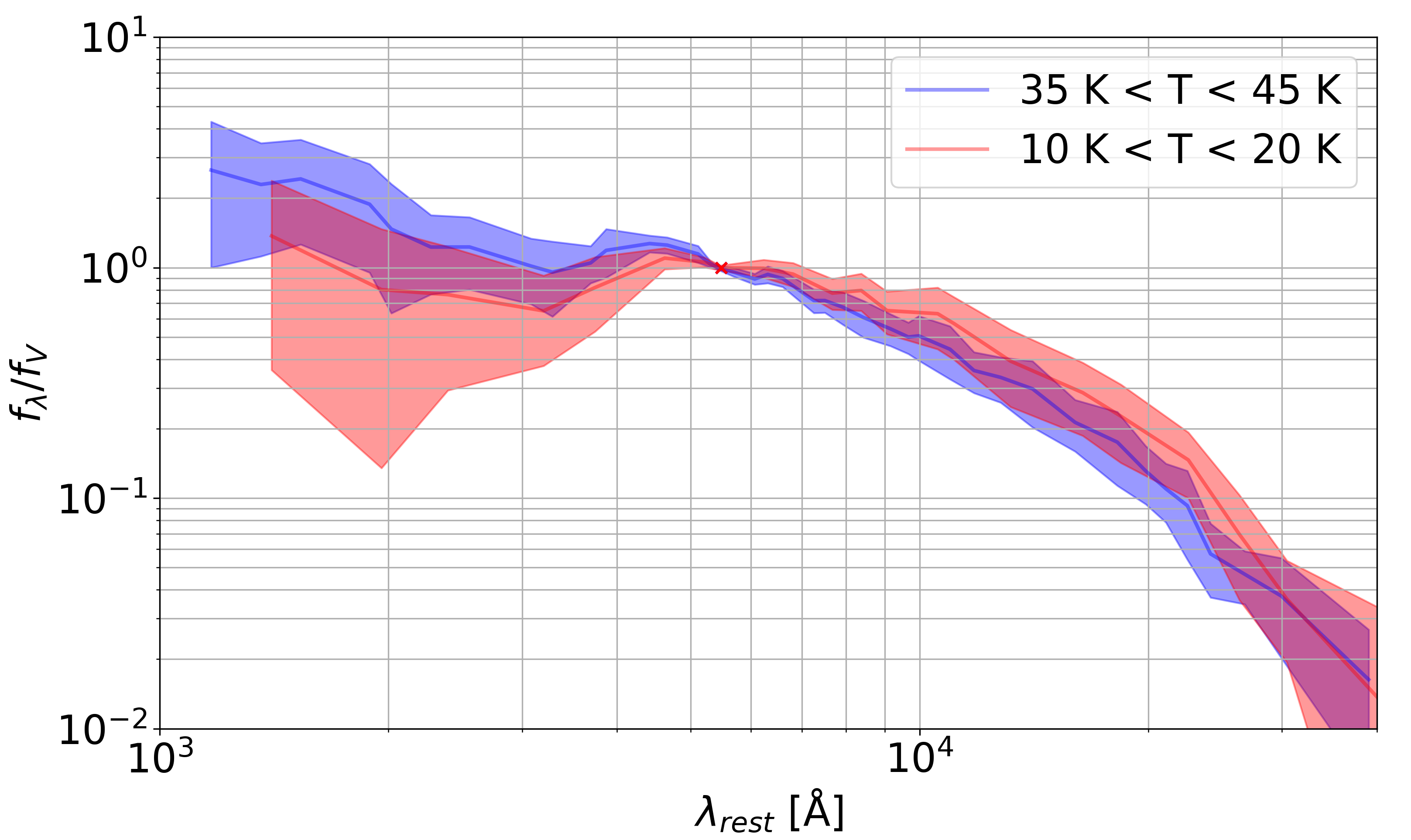}
\caption{Composite SED normalised at 5600 Å for galaxies with best-fit temperature in 10-20 K range [red] and 35-45 K range [blue] with standard deviation in shading. The central line indicates median SED with shading indicating 16 and 84 percentiles. The cooler galaxies display systematically less flux than the hotter galaxies at all wavelengths of the ultraviolet [$\lambda <$ 4000 Å]. Cooler galaxies look more quiescent than the hotter galaxies, which share spectral features with more star-forming galaxies.} 
\label{fig:stack}
\end{center}
\end{figure}

The wide range of best-fit stellar initial mass functions can also be validated by inspecting the spectral energy distributions. A composite observed SED has been constructed combining the COSMOS2015 photometry with known spectroscopic redshifts using the EAZY-py implementation for composite SED (https://github.com/gbrammer/eazy-py, \cite{Brammer2008}). The observed photometry is (using filter-profiles and the spectroscopic redshifts) converted to an estimated rest-frame flux. Thus, using the multitude of bands and the wide range of redshifts one can sample a galaxy-populations rest-frame spectrum. These observed SED are not corrected for dust attenuation, which is however included in the fitting procedure for $T_{IMF}$. As they do not account for dust, composite SED have limited interpretability, but are useful for determining if separations in parameters-space leads to a separation of objects in color-space.
The stacked photometry can be seen for galaxies split in two ranges of best-fit temperatures in Fig. \ref{fig:stack}. Importantly, these populations show different spectral shapes, with more flux in ultraviolet (UV) and bluer part of the optical spectrum for hotter galaxies. Thus, stratifying objects in $T_{IMF}$ leads to a separation of objects in the specific regions of color-space most sensitive to a change in IMF.



\begin{figure*}
    \centering
    \includegraphics[width=\linewidth]{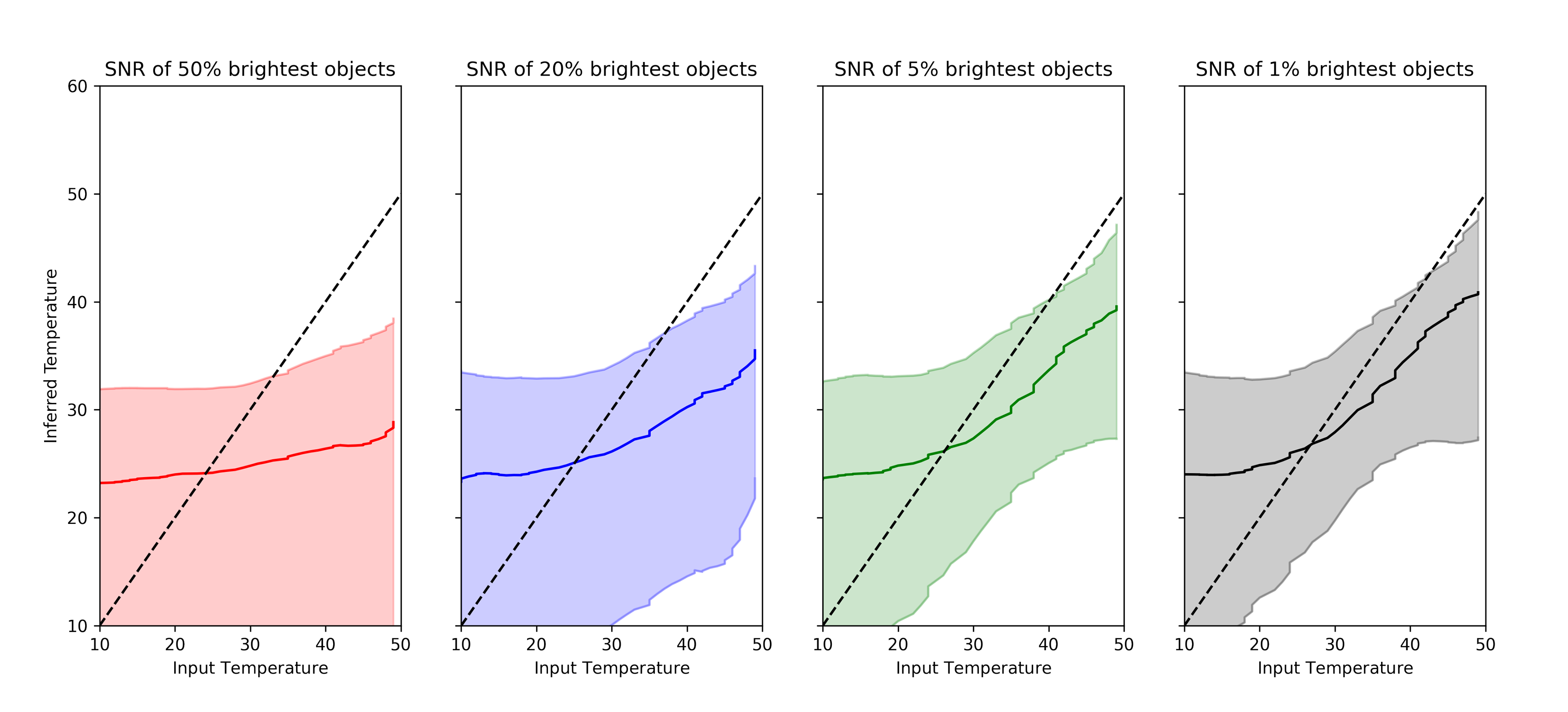}
    \caption{Median of inferred temperatures versus input temperatures of synthetic photometry of the EAZY template basis. Shading indicates lower and upper quartiles}. For each panel 10,000 synthetic photometry are generated with a known input temperatures ranging from 10 to 50 K. The photometry is scattered with Gaussian uncertainties in each filter. The uncertainties are drawn corresponding to the representative SNR for each band within COSMOS2015 [from left to right the panels show the SNR from the median, 80th, 95th and 99th percentile]. With increasing SNR, the inferred temperature increasingly matches the underlying input temperature. However, even for the 99th percentile a systematic bias between input and inferred exists, where especially bottom-heavier IMFs remain hard to distinguish.  
    \label{fig:Tinp_Toup}
\end{figure*}

\subsection{Quality Cuts}
\label{subsec:qualitycuts}

A final concern is the possibility that the tight relationship between best-fit $T_{IMF}$ and redshift arises from the cuts applied to the sample, rather than from a true astrophysical relationship. However, this concern may be alleviated by applying the same approach to synthetic photometry of spectra with known $T_{IMF}$. Here photometry has been constructed using the EAZY template basis, with scatter characteristic of galaxies at different signal-to-noise from the COSMOS2015 catalog.  A best-fit IMF temperature (Fig. \ref{fig:Tinp_Toup}) is then determined.  For the majority of objects in the COSMOS2015 catalog, the photometry is not constrained well enough to determine $T_{IMF}$, but for the highest-quality objects, the input $T_{IMF}$ can be recovered. 

The spread observed between input and inferred temperatures of synthetic photometry is typically about 10 K for the 95th percentile; This is significantly larger than the 5 K spread suggested by the uncertainty from the Monte Carlo simulations (see \S \ \ref{sec:sig}) and the physical spread observed at fixed redshift (see \S \ \ref{fig:temp-z-all}). This reflects a generic issue present in template-fitting codes when the observed galaxies are outside the synthetic models parameter-space. Any fitting would in such cases typically happen at the bound of the parameter-space, which would artificially suppress the variance \citep{Brammer2008}. In future, this could be alleviated with a deeper understanding of galactic spectra and improvements to FSPS modeling, but this analysis remains agnostic, since the $T_{IMF}$ for an ensemble of galaxies is highly significant with either metric of uncertainty.

\begin{figure}[ht]
\begin{center}
\includegraphics[width=\linewidth]{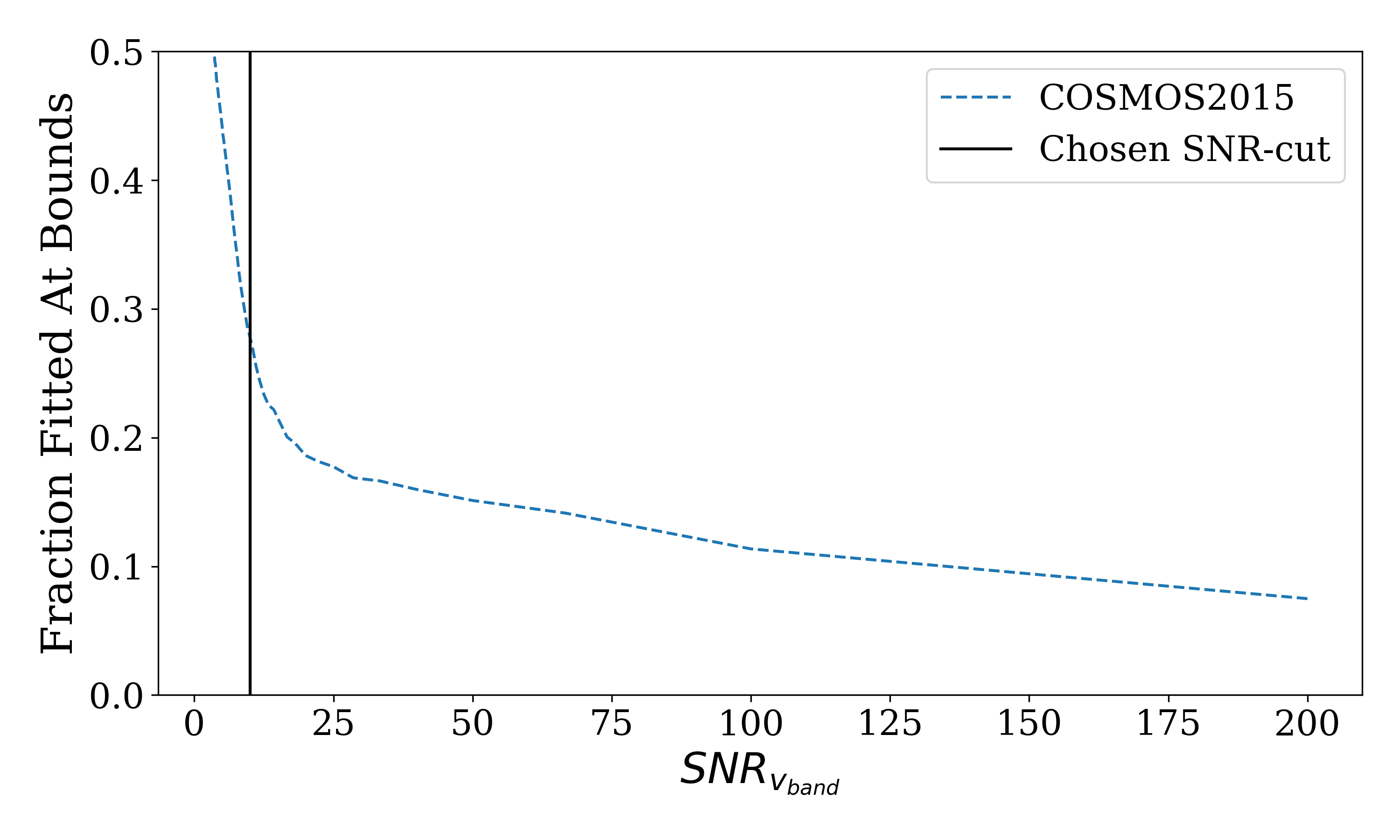}
\caption{Fraction of objects best fit at the bounds of temperature range as a function of SNR in $V_{band}$. As noise decreases the fraction of outliers decreases monotonically. Note in paper III, it is shown that a SNR-cut in $K_{band}$ yields equivalent results. Objects still best fit at the temperature bounds at high SNR typically have little coverage within UV and optical bands. } 
\label{fig:SNR_fraction}
\end{center}
\end{figure}

Additionally, the average best-fit $T_{IMF}$ is consistent with no systematic bias for increasing uncertainties on flux-measurements. However, the fraction of objects best fit at the bounds of the temperature range monotonically increases with increased noise (see Fig. \ref{fig:SNR_fraction}). Therefore, the well-measured galaxies are important for constraining the IMF, but not for the overall trend with redshift seen in \S~\ref{sec:results}. The inclusion of temperatures outliers does not influence the trends presented in this paper; The redshift evolution and the reinterpretation of galactic properties displays the distribution and dependence of $T_{IMF}$, while the differences between ULIRGs, star-forming and quiescent galaxies remains valid with or without outliers. Typically, objects fit at the bounds are ill-constrained because of limited coverage in optical and UV. Thus, because the inferred temperature may not represent the underlying IMF, we recommend caution when interpreting these outlying $T_{IMF}$.


\section{Discussion}
\label{sec:discussion}
This work introduces an additional parameter, $T_{IMF}$, into photometric template fitting.  This allows a selection of the stellar initial mass function (IMF) from a theoretical family of possible IMFs, with a higher $T_{IMF}$ producing a top-heavier stellar mass distribution.  A Milky Way-like IMF (here, a Kroupa IMF), which previously has typically been assumed to be Universal, is part of this family and corresponds to $T_{IMF} = 20$ K.

Fitting the COSMOS2015 catalog with this additional parameter produces three key results:
\begin{itemize} 
    \item {{\bf Top-Heavier IMFs}: Most galaxies are best-fit with $T_{IMF} > 20$ K, typically ranging from 25-40 K.  This means that most galaxies should have top-heavier stellar populations than previously expected.  One likely consequence is that stellar masses($M_*$) and star-formation rates (SFR) have previously been overestimated, with greater overestimation associated with galaxies with higher $T_{IMF}$.  However, due to covariances between $T_{IMF}$, extinction, and metallicity, this is not necessarily true for \emph{every galaxy} with $T_{IMF} > 20$ K.}  %
    \item {{\bf An IMF "main sequence"}: At any fixed redshift, star-forming galaxies span a much narrower range of $T_{IMF}$ than the full sample.  For $z>2.5$, where the vast majority of all galaxies in COSMOS2015 are star-forming, the standard deviation in $T_{IMF}$ is a low as 2-3 K.  At lower redshift, the star-forming galaxies alone have a comparable scatter.  In addition, the typical $T_{IMF}$ for star-forming galaxies increases towards higher redshift, from $\sim 25$ K locally to $\sim 35$ K by $z = 4$.  
    
    These two properties, a universality in $T_{IMF}$ at fixed redshift and a monotonic redshift dependence, have been seen in several other properties of evolving galaxies.  The star-forming "main sequence" finds a tight relationship between SFR and $M_*$ at fixed redshift, with SFR becoming more efficient towards high redshift \citep{Peng2010,Speagle2014,Renzini2015,Tomczak2016}.  Similar evolution, at least qualitatively, is found in dust temperatures \citep{Papadopoulos2010b,Casey2012,Magnelli2014,Magdis2017}.  Finally, quasars also exhibit a "main sequence"-like relationship between luminosity ($\dot{M}$, analogous to SFR) and virial black hole mass (analagous to $M_*$) \citep{Steinhardt2010a,Steinhardt2011}.  If $T_{IMF}$ indeed reflects the gas temperature in cool molecular clouds, it is natural to look to explain all of these simultaneously with the same strong feedback model \citep{Kelson2014,Steinhardt2014b,Peng2015}.}
    \item {{\bf Relationship Between IMF and Evolutionary State}: At low redshift, the COSMOS2015 includes a large sample of both star-forming and quiescent galaxies.  Star-forming galaxies exhibit higher $T_{IMF}$ than quiescent galaxies.  In the other direction, although $T_{IMF}$ for star-forming galaxies grows toward high redshift, $T_{IMF}$ for ULIRGs is both higher and consistent with no redshift evolution.  Thus, the IMF appears not just to evolve with redshift or time, but also to reflect the evolutionary state of its host galaxy.}
\end{itemize}

An additional conclusion is that if previous methods assuming a Galactic IMF have incorrectly estimated key properties, relationships such as the star-forming main sequence might need to be re-evaluated.  These are discussed in detail in the remaining two papers in this series: Paper II covers the star-forming population, and Paper III focuses on quiescent galaxies.

\subsection{Interpretation of  \texorpdfstring{$T_{IMF}$}{TIMF}}

According to the prescription in \citet{Jermyn2018}, $T_{IMF}$ measures the gas temperature in star-forming clouds.  Different studies have proposed a similar family of IMFs, but with a different temperature dependence.  The same photometric template fitting routine can be used for any of these models, but the relationship between gas temperature $T_g$ and $T_{IMF}$ will vary (see Table \ref{tab:tdependence} for the various conversions).  Any of these models would produce a $T_g$ higher than in the Milky Way.  However, the typical $T_g$ at $z \sim 4$, corresponding to $T_{IMF} = 35$ K, can range from 31 to 61 K.  

If gas and dust are in thermal equilibrium, as might be expected for sufficiently dense clouds, then it might be possible to use dust temperatures to properly calibrate the $T_g - T_{IMF}$ relation.  Dense clouds may also be more common sites of star formation, both because of higher gravitational potential and the possibility for more efficient cooling.  Dust temperatures ($T_d$) along the star-forming main sequence have been measured to increase from $\sim 25$ K locally to $\sim 32$ K by $z = 2$ \citep{Magnelli2014}.  This would be most consistent with the \citet{imf-t52} or \citet{Steinhardt2020} scaling relations.  The systematic uncertainty due to different dust temperature scaling relations is too large to distinguish between the two.
 
However, there are two significant differences between gas and dust temperatures. First, measured dust temperatures are typically luminosity-weighted across the entire galaxy. This signal is dominated by luminous (and thus hotter) clouds leading to an overestimation of the average $T_d$.  $T_g$ as determined by the IMF, on the other hand, will be weighted by star formation, and therefore likely dominated by the coolest large clouds.  Second, $T_d$ is measured using radiation from dust, which means it provides a temperature at the time when those photons were emitted.  However, $T_g$ is measured using the IMF responsible for the bulk of the existing stellar population.  For an old stellar population, this may be significantly earlier than the time photons were emitted.  Thus, even assuming thermal equilibrium, $T_g$ and $T_d$ might diverge.  

An additional effect is that although $T_{IMF}$, and thus $T_g$, is observed to drop towards low redshift for star-forming galaxies, this cannot continue into quiescence.  Once a galaxy becomes quiescent, the bulk of the stellar population should consist of the same stars (absent the very high-mass end).  Since $T_{IMF}$ measures the conditions under which those stars formed, it must be constant as well.  The possibility that the evolution of $T_g$ could be used to select quenching galaxies is discussed in more detail in Paper III.

\subsection{Limitations From Assuming a Monolithic IMF}

In order to introduce only a single additional parameter, in this work galaxies are modeled as having a single IMF for their entire stellar population.  This assumption is very likely unphysical.  If $T_{IMF}$ evolves with redshift, as found here, then the stellar population in a galaxy would be comprised of components produced from different IMFs.  Further, unless the dominant contribution to galactic heating truly permeates the entire galaxy, different molecular clouds should have different gas temperatures.  This issue might be less critical, since the coolest clouds should be the sites of the bulk of the star formation.  

In some respects, the same is true of other parameters in photometric template fitting.  Galaxies in the COSMOS2015 catalog are described as having a single, luminosity-weighted, age, extinction, and metallicity (or, in some cases, probability distributions for that single parameter; \citealt{Laigle2016}).  Fitting portions of galaxies which can be resolved into multiple components often results in these parameters varying between those components \citep{Greener2020,Fetherolf2020}.  

EAZY fits galaxy photometry as a linear combination of a basis which includes components of different age, extinction, and metallicity.  Thus, at least in principle, a galaxy with distinct subpopulations could be well-fit with this technique.  However, the 12 basis vectors are insufficient to span the full space of subpopulations.  

Introducing the additional parameter of $T_{IMF}$ would not have allowed 12 eigenspectra to span the full space spanned by the standard EAZY template basis.  Expanding the size of the basis would have required too many additional measurements to constrain the best fit, so that high-redshift objects could not be modeled.  As a result, instead $T_{IMF}$ is treated differently than these other parameters, fit from a grid of single values.  A grid-like approach has been used successfully in other template-fitting codes, which model the full stellar population with a single set of parameters.  However, this means that stellar populations produced from multiple IMFs will be fit with a single, luminosity-averaged, IMF.  As a galaxy becomes quiescent, even without new star formation the best-fit $T_{IMF}$ could vary, as the luminosity-averaged IMF transitions from dominated by the youngest population to older ones.  

\subsection{Adding Additional Parameters}

One solution would be adding a second parameter, corresponding to an evolutionary history of the IMF.  It would be straightforward to this parameter to span a reasonable range of shapes for that history.  However, it should be noted that the addition of one parameter required a significantly more stringent quality cut (Fig. \ref{fig:qualitycut}) than in the full COSMOS2015 catalog.  At present, only a small fraction of objects in COSMOS2015 can be well-modeled, and other photometric surveys typically have fewer available bands.  Adding a second new parameter would have a similar effect, and it might become impossible to fit any objects at, e.g., $z > 2$.

If sufficient information is available to add a parameter, perhaps a more useful choice would be the dust extinction curve.  As with the IMF, currently a single extinction curve is typically assumed for fitting an entire photometric catalog, with only the normalization determined by the fit.  However, there is a strong argument for variability with measurements of local galaxies finding a range of different extinction curves \citep{Gordon2003,Salim2020}.  

There is also the possibility of a strong covariance between the best-fit extinction curve and the best-fit IMF shape.  Thus, modifying template fitting codes to investigate this possibility would provide an important test of the robustness of the results in this work.  This likely requires a substantial rewrite to either EAZY or a different template-fitting code, and is thus beyond the scope of this work.

Another significant parameter not currently included is the shape of the star-formation history (SFH).  Although there is theoretically a complete degeneracy between the SFH and the IMF, in practice different reasonable choices of SFH produced similar $T_{IMF}$ (Fig. \ref{fig:SFH-comp}). 


Ultimately, despite the limitations of adding only a single parameter to describe the range of possible stellar initial mass functions, the best-fit IMF is well constrained from $\sim 30$-band, multi-wavelength photometry.  For most galaxies, this produces an IMF different than that of our own Galaxy.  Because nearly every quantity derived from photometric template fitting is sensitive to the IMF, finding ways to include a range of IMFs should be considered necessary for future template fitting pipelines.  The resulting implications for models of star-forming galaxies (Paper II) and quiescent galaxies (Paper III) are discussed in companion papers.    

\acknowledgements

The authors would like to thank Gabe Brammer, Tommy Clark, Andrei Diaconu, Vasily Kokorev, Vadim Rusakov, Sune Toft, and Darach Watson for helpful comments.  CLS is supported by ERC grant 648179 ``ConTExt''.  The Cosmic Dawn Center (DAWN) is funded by the Danish National Research Foundation under grant No. 140.  The Flatiron Institute is supported by the Simons Foundation.


\bibliographystyle{mnras}
\bibliography{refs.bib}

\label{lastpage}
\end{document}